\documentclass[preprint,12pt]{elsarticle}

\usepackage{amssymb}

\usepackage[dvipsnames]{xcolor}
\usepackage[utf8]{inputenc}
\usepackage[T1]{fontenc}
\usepackage{array,multirow}
\usepackage{comment}

\usepackage{natbib}
\biboptions{sort&compress}

\usepackage{lineno}
\usepackage[
  letterpaper,
  left=2.5cm,
  right=2.5cm
]{geometry}

\usepackage{amsmath}

\journal{???}

\newcommand\Deb{\mbox{\textit{De}}} % Deborah number
\newcommand\Wei{\mbox{\textit{Wi}}} % Weissenberg number
\newcommand\Oh{\mbox{\textit{Oh}}}  % Ohnesorge number
\newcommand\Bo{\mbox{\textit{Bo}}}  % Bond number
\begin{document}

\begin{frontmatter}



\title{When does the elastic regime begin in viscoelastic pinch-off?}

%Ventilation optimization concepts for energy saving and aerosol safety


\cortext[cor]{Corresponding author}

\author[aff1]{A. Gaillard}
\ead{antoine0gaillard@gmail.com}
\author[aff2]{M. A. Herrada}
\author[aff1]{A. Deblais}
\author[aff1]{C. van Poelgeest}
\author[aff1]{L. Laruelle}
\author[aff3]{J. Eggers}
\author[aff1]{D. Bonn}

\affiliation[aff1]{organization={Van der Waals-Zeeman Institute, University of Amsterdam}, %Department and Organization
            addressline={Science Park 904}, 
            city={Amsterdam},
            %postcode={}, 
            %state={},
            country={Netherlands}}
            
\affiliation[aff2]{organization={Depto. de Mecánica de Fluidos e Ingeniería Aeroespacial, Universidad de Sevilla}, %Department and Organization
            %addressline={E-41092}, 
            city={Sevilla},
            postcode={E-41092}, 
            %state={},
            country={Spain}}

\affiliation[aff3]{organization={School of Mathematics, University of Bristol}, %Department and Organization
            addressline={University Walk}, 
            city={Bristol},
            postcode={BS8 1 TW}, 
            %state={},
            country={United Kingdom}}

\begin{abstract}
In this experimental and numerical study, we revisit the question of the onset of the elastic regime in viscoelastic pinch-off. This is relevant for all modern filament thinning techniques which aim at measuring the extensional properties of low-viscosity polymer solutions such as the Slow Retraction Method (SRM) in Capillary Breakup Extensional Rheometry (CaBER) as well as the dripping method where a drop detaches from a nozzle. In these techniques, a stable liquid bridge is slowly brought to its stability threshold where capillary-driven thinning starts, slowing down dramatically at a critical radius $h_1$ marking the onset of the elastic regime where the bridge becomes a filament with elasto-capillary thinning dynamics. While a theoretical scaling for this transition radius exists for the classical step-strain CaBER protocol, where polymer chains stretch without relaxing during the fast plate separation, we show that it is not necessarily valid for a slow protocol such as in SRM since polymer chains only start stretching (beyond their equilibrium coiled configuration) when the bridge thinning rate becomes comparable to the inverse of their relaxation time. We derive a universal scaling for $h_1$ valid for both low and high-viscosity polymer solution which is validated by both CaBER (SRM) experiments with different polymer solutions, plate diameters and sample volumes and by numerical simulations using the FENE-P model. 

\end{abstract}

%  \centerline{\includegraphics[scale=0.5]{figures/graphical_abstract}}

%\item Highlight 1

\begin{keyword}
Viscoelasticity \sep Filament \sep Pinch-off 




\end{keyword}

\end{frontmatter}

%\tableofcontents


%\newpage
%%%%%%%%%%%%%%%%%%%%%%%%%%%%%%%%%%%%%%%%%%%%%%%%%%%%%%%%%%%%%%%%%%%%%%%%%%%%%%%%%%%%%%%%%%%%%%%%%%%%%%%%%%%%%%%%%%%%%%%%%%%%%%%%
\section{Introduction}
\label{sec:Introduction}

The elasticity of a polymer solution can be probed by stretching a drop between one's thumb and index finger, resulting in the formation of a filament with a persistence time which is linked to the relaxation time of the solution. Such filaments are observable in many industrial free-surface flows such as spraying \cite{gaillard2022determines,keshavarz2015studying,keshavarz2016ligament} and inkjet printing \cite{sen2021retraction,christanti2002effect}, where long polymer molecules can be added to a Newtonian solvent to achieve a specific flow property, as well as in ejecta produced when coughing and sneezing \cite{scharfman2016visualization,gidreta2023effects}. The capillary-driven thinning dynamics of these filaments is the basis of numerous rheometry techniques dedicated to low-viscosity fluids, for which other techniques such as Meissner's RME (Rheometric Melt Elongation rheometer) and FiSER (Filament Stretching Extensional Rheometer) are not applicable. These techniques include Capillary Breakup Extensional Rheometry (CaBER) where a droplet is confined between two plates which are separated beyond the the range of stable liquid bridges \cite{bazilevsky1997failure,anna2001elasto,stelter2000validation}, the dripping technique where a droplet detaches from a nozzle \cite{Amarouchene,tirtaatmadja2006drop,rajesh2022transition} and Dripping-onto-Substrate (DoS) where a solid substrate is brought into contact with a drop hanging steadily from a nozzle \cite{dinic2017pinch}. All these techniques aim at creating a viscoelastic filament by triggering the pinching of a liquid column via the Rayleigh-Plateau instability.

Viscoelastic filaments are found to thin exponentially over time for a wide range of polymer-solvent systems and polymer concentrations (dilute and semi-dilute), consistent with the Oldroyd-B model which predicts
\begin{equation}
    h = h_1 \exp{\left( -\frac{t-t_1}{3\tau} \right)},
\label{eq:exponential}
\end{equation}
where $h$ is the (minimum) filament radius and $\tau$ the relaxation time of the polymer solution, the longest one for a multimode model \cite{anna2001elasto,entov1997effect}. This regime corresponds to an elasto-capillary balance where the elastic stress arising from the stretching of polymer chains balances the driving capillary pressure. Experimentally, starting from an equilibrium situations where polymers are relaxed (no pre-stress), this elastic regime can only be observed once polymers have been sufficiently stretched to overcome inertia and/or viscosity, which occurs at a time $t_1$ and at a filament radius $h_1 = h(t_1)$ marked by a sudden deceleration of the thinning dynamics. 

The amount of stretching of polymer chains at times $t<t_1$ is set by the strength of the extensional flow in the pinching region. In the limit case where the thinning dynamics at times $t<t_1$ (before elasticity balances capillarity) is much faster than the solution's relaxation time, i.e., where polymer chains deform by the same amount as the surrounding solvent itself without relaxing, Clasen et al. \cite{clasen2006beads} showed that the Oldroyd-B model leads to
\begin{equation}
    h_1 = \left( \frac{G h_0^4}{2 \gamma} \right)^{1/3},
\label{eq:h1_norelaxation}
\end{equation}
where $\gamma$ is the surface tension, $G$ the elastic modulus and $h_0$ the radius of the `initial' liquid column before the onset of capillary thinning, i.e., when the fluid is still at rest. This formula was first derived by Bazilevsky et al. \cite{bazilevsky1997failure} and differs by a factor $2^{1/3}$ from the formula proposed by Entov and Hinch \cite{entov1997effect} who did not treat the tension in the filament properly. 

This `relaxation-free' scenario leading to equation \ref{eq:h1_norelaxation} corresponds to the step-strain CaBER protocol where the plates are separated so fast that polymer chains stretch without having time to relax as the liquid bridge, connecting the two plates, stretches axially. In this step-strain protocol, the plates are separated exponentially over time to create an extensional flow with a constant extension rate $\dot{\epsilon}_0$ which, to ensure that polymer relaxation is negligible, must be larger that the coil-stretch transition value $1/2\tau$ \cite{miller2009effect}. This corresponds to Weissenberg number $\Wei_0 = \dot{\epsilon}_0 \tau > 1/2$. Once the plates have reached their final separation distance $L_f$, the unstable liquid bridge between the two plates continues to thin, this time under the action of capillarity, until the elastic regime starts at a bridge / filament (minimum) radius $h_1$. Miller et al \cite{miller2009effect} showed that, consistent with equation \ref{eq:h1_norelaxation}, $h_1$ does not depend on $L_f$ for polymer solutions. We are however unaware of experimental studies where $h_1$ was reported and tested against equation \ref{eq:h1_norelaxation} for different plate diameters and initial gaps, which set the radius $h_0$ of the initial (unloaded) fluid sample, or for different polymer solutions.

This step-strain CaBER protocol is however not recommended for low-viscosity polymer solutions since a fast plate separation leads to inertio-capillary oscillations of the end-drops which hinder the measurement of the relaxation time \cite{rodd2005capillary}. Alternative protocols consist in reaching the threshold of the Rayleigh-Plateau instability slowly, e.g. by separating the plates at a constant low velocity in CaBER (Slow Retraction Method or SRM) \cite{campo2010slow}. In that case, the initially stable liquid bridge connecting the two end-plates becomes unstable at a critical plate separation distance, corresponding to a minimum bridge radius $h_0$, and thins further under the action of capillarity. This is similar to the dripping method where the bridge connecting a droplet to a nozzle, from which liquid is infused at a low flow rate, becomes unstable at a critical droplet weight \cite{rajesh2022transition}. 

In such slow protocols, equation \ref{eq:h1_norelaxation} may not be valid if the time taken by the bridge to thin from its initial (minimum) radius $h_0$ to the radius $h_1$ (marking the onset of the elastic regime) is larger than the liquid's relaxation time $\tau$, as was already noticed by Bazilevsky et al. \cite{bazilevsky1997failure}. In that case, polymer chains may indeed remain in a coiled state for a significant time, only starting to stretch when the bridge's thinning rate becomes comparable to $1/\tau$. This led Campo-Deaño \& Clasen \cite{campo2010slow} to derive an alternative formula for $h_1$ for their slow retraction CaBER method which, to our knowledge, has never been tested experimentally. In this formula, $h_1$ is independent of $h_0$, in sharp contrast with equation \ref{eq:h1_norelaxation} which predicts $h_1 \propto h_0^{4/3}$. In a more recent experimental work from Rajesh et al. \cite{rajesh2022transition}, the authors proposed an empirical scaling $h_1 \propto R_n^{0.66}$ in dripping experiments with low-viscosity polymer solutions where $R_n$ in the nozzle radius, but they did not provide a theoretical explanation for their findings.

In such slow protocols, equation \ref{eq:h1_norelaxation} is only expected to be valid if the time taken by the liquid bridge to thin from $h_0$ to $h_1$ is much shorter than the liquid's relaxation time, in which case polymer chains stretch without having time to relax. This time is expected to scale as the characteristic time scale of the capillary-driven bridge thinning dynamics derived from linear stability theory, namely, the Rayleigh (inertio-capillary) time scale \cite{wagner2005droplet}
\begin{equation}
\tau_R = (\rho h_0^3 / \gamma)^{1/2},
\label{eq:tau_R}
\end{equation}
\noindent or the visco-capillary time scale 
\begin{equation}
\tau_{\mathrm{visc}} = \eta_0 h_0/\gamma,
\label{eq:tau_visc}
\end{equation}
\noindent depending on the Ohnesorge number 
\begin{equation}
\Oh = \frac{\eta_0}{\sqrt{\rho \gamma h_0}} = \frac{\tau_{\mathrm{visc}}}{\tau_R},
\label{eq:Oh}
\end{equation}
\noindent where $\rho$ and $\eta_0$ are the liquid density and total (zero-shear) viscosity, respectively. In other words, if we define a Deborah number 
\begin{equation}
\Deb = \tau/\tau_R
\label{eq:De}
\end{equation}
\noindent based on the Rayleigh time scale, equation \ref{eq:h1_norelaxation} is expected to be valid for $\Deb \gg 1$ in the inviscid case ($\Oh \ll 1$) and for $\Deb / \Oh = \tau / \tau_{\mathrm{visc}} \gg 1$ in the viscous case ($\Oh \gg 1$), which is the limit considered in most analytical studies \cite{clasen2006beads}.

In this study, we aim to expand our current understanding of the transition radius $h_1$ (marking the onset of the elastic regime) to cases where polymer relaxation is not negligible during the capillary driven thinning of the liquid bridge. This discussion follows up on our previous paper where $h_1$ was observed to increase linearly with $h_0$ for different liquids for a slow plate separation CaBER protocol \cite{gaillard2023beware}, a scaling which differs from the $h_1 \propto h_0^{4/3}$ prediction of equation \ref{eq:h1_norelaxation}. Materials and methods are presented in \S\ref{sec:materials} and experimental results are presented in \S\ref{sec:Experimental results on $h_1$}. Theoretical expressions of $h_1$ are derived and tested experimentally and numerically using the Oldroyd-B model in \S\ref{sec:Oldroyd-B prediction for $h_1$} and the FENE-P model in \S\ref{sec:FENE-P prediction for $h_1$}.

%%%%%%%%%%%%%%%%%%%%%%%%%%%%%%%%%%%%%%%%%%%%%%%%%%%%%%%%%%%%%%%%%%%%%%%%%%%%%%%%%%%%%%%%%%%%%%%%%%%%%%%%%%%%%%%%%%%%%%%%%%%%%%%%
\section{Materials and methods}
\label{sec:materials}

The liquids, their shear rheology and the setup used in experiments are presented in \S\ref{subsec:Liquids} and \S\ref{subsec:Shear rheology} and \S\ref{subsec:CaBER setup and protocol}, respectively, while the equations and numerical methods used in simulations are presented in \S\ref{subsec:Numerical methods}.

\subsection{Liquids}
\label{subsec:Liquids}

\noindent Three of the polymer solutions used in the present study are the same as in our previous paper \cite{gaillard2023beware} which have comparable `relaxation times' or, more precisely, comparable filament thinning rates. Two of them are solutions of poly(ethylene oxide) (PEO) of molecular weight $M_w = 4 \times 10^{6}$~g/mol (PEO-4M), one in water with concentration $500$~ppm, referred to as PEO$_{\mathrm{aq}}$, and one in a $\sim 260$ times more viscous solvent with concentration $25$~ppm, referred to as PEO$_{\mathrm{visc}}$, and the third one is a $1000$~ppm solution of poly(acrylamide/sodium acrylate) (HPAM) [70:30] of molecular weight $M_w = 18 \times 10^{6}$~g/mol in water with $1$~wt\% NaCl to screen electrostatic interactions and make the chain flexible instead of semi-rigid. Both polymers were provided by Polysciences (ref. 04030-500 for PEO and 18522-100 for HPAM). The solvent of the PEO$_\mathrm{visc}$ solution is a Newtonian $30$~wt\% aqueous solution of poly(ethylene glycol) (PEG) with molecular weight $20,000$~g/mol (PEG-20K). After slowly injecting the polymer powder to a vortex generated by a magnetic stirrer, solutions were homogenised using a mechanical stirrer at low rotation speed for about $16$ hours. For the PEO$_{\mathrm{visc}}$ solution, PEG was added after mixing PEO with water. Additional solutions of PEO-4M in water were prepared from dilution of a $10,000$~ppm stock solution with concentrations ranging between $5$ and $10,000$~ppm to investigate the influence of polymer concentration.

\subsection{Shear rheology}
\label{subsec:Shear rheology}

\begin{figure}
  \centerline{\includegraphics[scale=0.58]{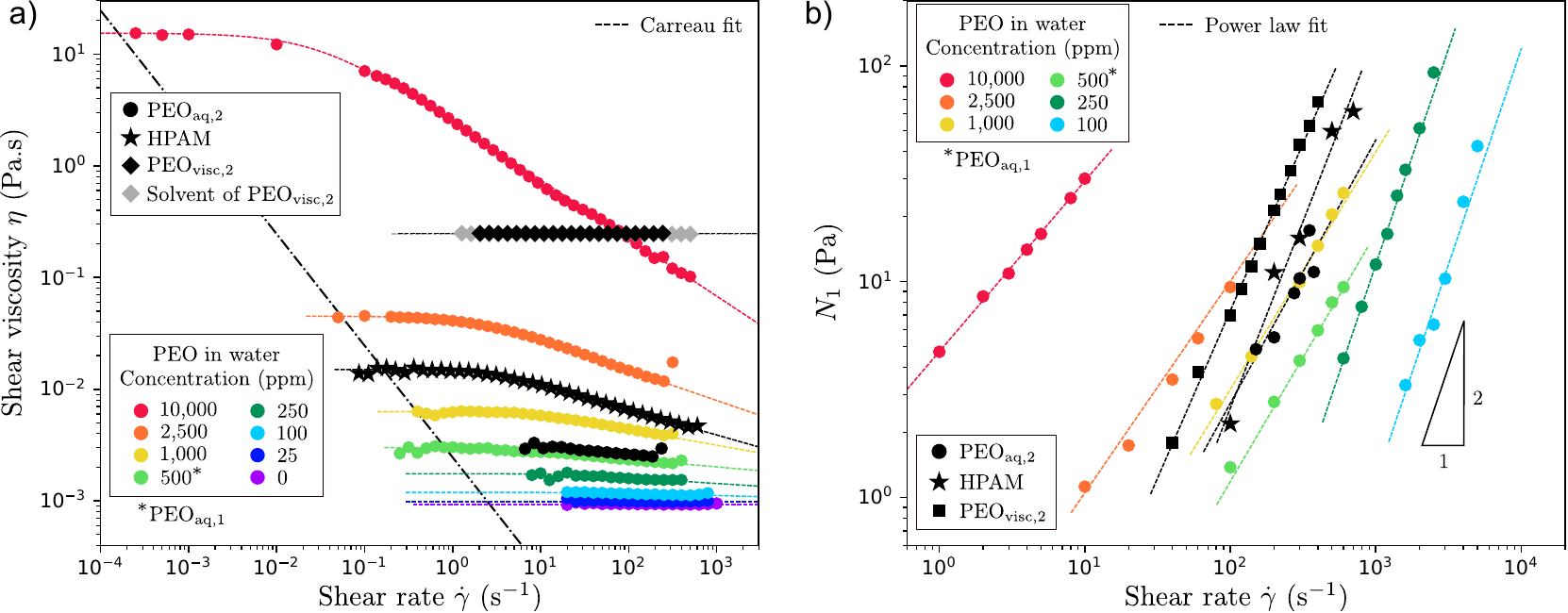}}
  \caption{(a) Shear viscosity $\eta$ and (b) first normal stress difference $N_1$ of the different polymer solutions against the shear rate $\dot{\gamma}$.}
\label{fig:shear_rheology}
\end{figure}

\begin{table}[t!]
\setlength{\tabcolsep}{10pt}
  \begin{center}
  \begin{footnotesize} % Only with this template.
\def~{\hphantom{0}}
  \begin{tabular}{cccccccccccc}
  
     $c$     & $\gamma$ & $c/c^*$ & $\eta_0$ & $\eta_p$ & $n$  & $1/\dot{\gamma}_c$ & $\alpha_1$ & $\Psi_1$ \\
    (ppm)    & (mN/m)   &         & (mPa~s)  & (mPa~s)  &      & (s)              &            & (Pa~s$^{\alpha_1}$)  \\ [8pt]      
     5       & 72.0     & 0.019   & 0.93     & 0.013    & 1    & --               & --         & -- \\
     10      & 72.0     & 0.037   & 0.940    & 0.02     & 1    & --               & --         & -- \\
     25      & 63.4     & 0.093   & 0.985    & 0.065    & 1    & --               & --         & -- \\
     50      & 62.8     & 0.19    & 1.04     & 0.12     & 1    & --               & --         & -- \\
     100     & 63.0     & 0.37    & 1.19     & 0.27     & 0.98 & 0.023            & 2          & $1.2 \times 10^{-6}$ \\
     250     & 63.0     & 0.93    & 1.75     & 0.83     & 0.95 & 0.054            & 2          & $1.2 \times 10^{-5}$ \\
     500     & 62.5     & 1.9     & 3.00     & 2.08     & 0.95 & 0.12             & 1.15       & $5.9 \times 10^{-3}$ \\
     1000    & 62.5     & 3.7     & 6.3      & 5.38     & 0.86 & 0.14             & 1.10       & $2.0 \times 10^{-2}$ \\
     2500    & 62.5     & 9.3     & 45       & 44       & 0.73 & 0.62             & 0.98       & $1.1 \times 10^{-1}$ \\
     10000   & 62.3     & 37      & 15400    & 15400    & 0.48 & 34               & 0.79       & $4.7 \times 10^{0}$ \\
  \end{tabular}
  \caption{Concentration $c$, reduced concentration $c/c^*$, surface tension $\gamma$ and shear rheological properties (from equations \ref{eq:carreau} and \ref{eq:N1}) of aqueous PEO-4M solutions prepared from dilution of the same $10,000$~ppm stock solution. $\eta_p = \eta_0-\eta_s$ is the polymer contribution to the shear viscosity. The density and solvent viscosity are $\rho = 998$~kg/m$^3$ and $\eta_s = 0.92$~mPa~s. The 500ppm solution in this table is referred to as PEO$_{\mathrm{aq,1}}$ in the text. For the $5$~ppm solution, $\eta_0$ is too close to $\eta_s$ to estimate $\eta_p$ and we therefore use $\eta_p = \eta_s [\eta] c$ with the intrinsic viscosity $[\eta]$ extracted from the linear fit of $\eta_p (c)$ for $c<c^*$.} 
  \label{tab:rheology_PEO}
  \end{footnotesize} 
  \end{center}
\end{table}

The shear viscosity $\eta$ and first normal stress difference $N_1$ of polymer solutions were measured at the temperature of CaBER experiments, typically $20^{\circ}$C, with a MRC-302 rheometer from Anton Paar equipped with a cone plate geometry (diameter $50$~mm, angle $1^{\circ}$ and truncation gap $53$~$\mu$m) and are shown in figure \ref{fig:shear_rheology}. To measure $N_1$, we follow a step-by-step protocol similar to Casanellas et al. \cite{casanellas2016stabilizing} in order to circumvent the instrumental drift of the normal force. This protocol consists in applying steps of constant shear rate followed by steps of zero shear and subtracting the two raw $N_1$ plateau values. The contribution of inertia to the normal force is corrected for by the rheometer \cite{macosko1994rheology}. We find that the PEO$_{\mathrm{visc}}$ solution is a Boger fluid with a constant shear viscosity while the HPAM solution is shear-thinning, as well as the aqueous PEO solutions when concentrations are larger than $250$~ppm. For shear-thinning solutions, the shear viscosity is fitted with the Carreau-Yasuda formula
\begin{equation}
\eta(\dot{\gamma}) = \eta_0 ( 1 + ( \dot{\gamma}/\dot{\gamma}_c)^{a_1})^{(n-1)/a_1},
\label{eq:carreau}
\end{equation}
\noindent where $\eta_0$ is the zero-shear viscosity, $n$ is the shear-thinning exponent and $\dot{\gamma}_c$ is the shear rate marking the onset of shear thinning, $a_1$ (typically $2$) encoding the sharpness of the transition towards the shear thinning regime. The polymer contribution to the shear viscosity $\eta_p = \eta_0 - \eta_s$ increases linearly with polymer concentration $c$ in the dilute regime and follows $\eta_p = \eta_s [\eta] c$ where, for the PEO solutions in water, we find an intrinsic viscosity $[\eta] = 2.87$~m$^3$/kg. Using the expression of Graessley \cite{graessley1980polymer}, gives a critical overlap concentration $c^* = 0.77/[\eta] = 0.268$~kg/m$^3$ ($268$~ppm), consistent with the onset of shear-thinning expected at $c>c^*$. For the PEO$_{\mathrm{visc}}$ solution, where only one concentration ($25$~ppm) was tested, assuming that the solution is dilute to calculate $[\eta]$ and $c^*$ using the same formulas leads to a larger critical overlap concentration  $c^* = 1400$~ppm, probably due to differences in polymer-solvent interactions (PEO in water vs. PEO in PEG solution). The first normal stress difference is fitted by a power law
\begin{equation}
N_1 = \Psi_1 \dot{\gamma}^{\alpha_1},
\label{eq:N1}
\end{equation}
\noindent where we find $\alpha_1 = 2$ below $c^*$ and $\alpha_1 < 2$ above $c^*$ for aqueous PEO solutions, and $\alpha_1 < 2$ for the PEO$_{\mathrm{visc}}$ and HPAM solutions. All fitting parameters are reported in table \ref{tab:rheology_PEO} for the PEO solutions of different concentrations in water, and in table \ref{tab:rheology_bigpins} for the PEO$_{\mathrm{aq}}$, PEO$_{\mathrm{visc}}$ and HPAM solutions. We also report the density $\rho$ and surface tension $\gamma$ measured with a pendant drop method and, when known, the ratio $c/c^*$.

\begin{table}
\setlength{\tabcolsep}{2.5pt}
  \begin{center}
  \begin{footnotesize} % Only with this template.
\def~{\hphantom{0}}
  \begin{tabular}{ccccccccccccc}
  
    Name~~                 & $\rho$   & $\gamma$ & $\eta_s$ & $c$   & $c/c^*$ & $\eta_0$ & $\eta_p$ & $n$  & $1/\dot{\gamma}_c$ & $\alpha_1$ & $\Psi_1$ & $\tau_{m}$ \\
                           &(kg/m$^3$)& (mN/m)   & (mPa~s)  & (ppm) &        & (mPa~s)   & (mPa~s)  &      & (ms)               &            & (Pa~s$^{\alpha_1}$) & (ms) \\ [8pt]      
    PEO$_{\mathrm{aq,2}}$~ & 998      & 62.5     & 0.92     & 500   & 1.9    & 3.3      & 2.08 & 0.93 & 120  &1.2 &$9.9 \times 10^{-3}$ & 240 \\
    PEO$_{\mathrm{visc,2}}$& 1048     & 56.0     & 245      & 25    & 0.018  & 248      & 3.3  & 1    & --   &1.6 &$5.8 \times 10^{-3}$& 110 \\
    HPAM~                  & 998      & 72.0     & 0.92     & 1000  & --     & 15       & 14   & 0.78 & 410  &1.7 &$9.0 \times 10^{-3}$& 100 \\
  \end{tabular}
  \caption{Properties of the polymer solutions used for plate diameters $2R_0$ up to 25~mm in CaBER measurements. $\rho$ is the density and $\gamma$ is the surface tension. See caption of table \ref{tab:rheology_PEO} for the definition of the shear properties. $\tau_{m}$ is the maximum CaBER relaxation time measured for the largest plates, see figure \ref{fig:h12_taue}(a). The PEO$_{\mathrm{visc,1}}$ and PEO$_{\mathrm{visc,2}}$ solutions have the same shear viscosity to within less than $5$~\%.} 
  \label{tab:rheology_bigpins}
  \end{footnotesize} 
  \end{center}
\end{table}

We must mention here that two different PEO$_{\mathrm{aq}}$ solutions and two different PEO$_{\mathrm{visc}}$ solutions have been used in this study, with differences in rheological properties in each case, caused by slightly different preparation protocols for a given recipe (e.g., a slightly different agitation time). The PEO$_{\mathrm{aq,1}}$ solution is prepared from dilution of the same stock solution as the other aqueous PEO solutions in table \ref{tab:rheology_PEO}. The PEO$_{\mathrm{aq,2}}$ solution featured in table \ref{tab:rheology_bigpins} exhibits a $10$\% larger shear viscosity and about $2.5$ times larger values of $N_1$, as shown in figure \ref{fig:shear_rheology}. The PEO$_{\mathrm{visc,1}}$ and PEO$_{\mathrm{visc,2}}$ solutions have the same shear viscosity to within less than $5$~\% and only the latter one is presented in figure \ref{fig:shear_rheology} and in table \ref{tab:rheology_bigpins}. As explained in \S\ref{subsec:CaBER setup and protocol}, the PEO$_{\mathrm{aq,1}}$ and PEO$_{\mathrm{visc,1}}$ solutions were tested with (CaBER) plate diameters less than 7~mm, varying the (non-dimensional) drop volume for each plate, whereas the PEO$_{\mathrm{aq,2}}$ and PEO$_{\mathrm{visc,2}}$ solutions were used for plate diameters up to 25~mm with a single (non-dimensional) drop volume for each plate.

\subsection{Experimental setup and slow-stepwise CaBER protocol}
\label{subsec:CaBER setup and protocol}

\begin{figure}[!ht]
  \centerline{\includegraphics[scale=0.57]{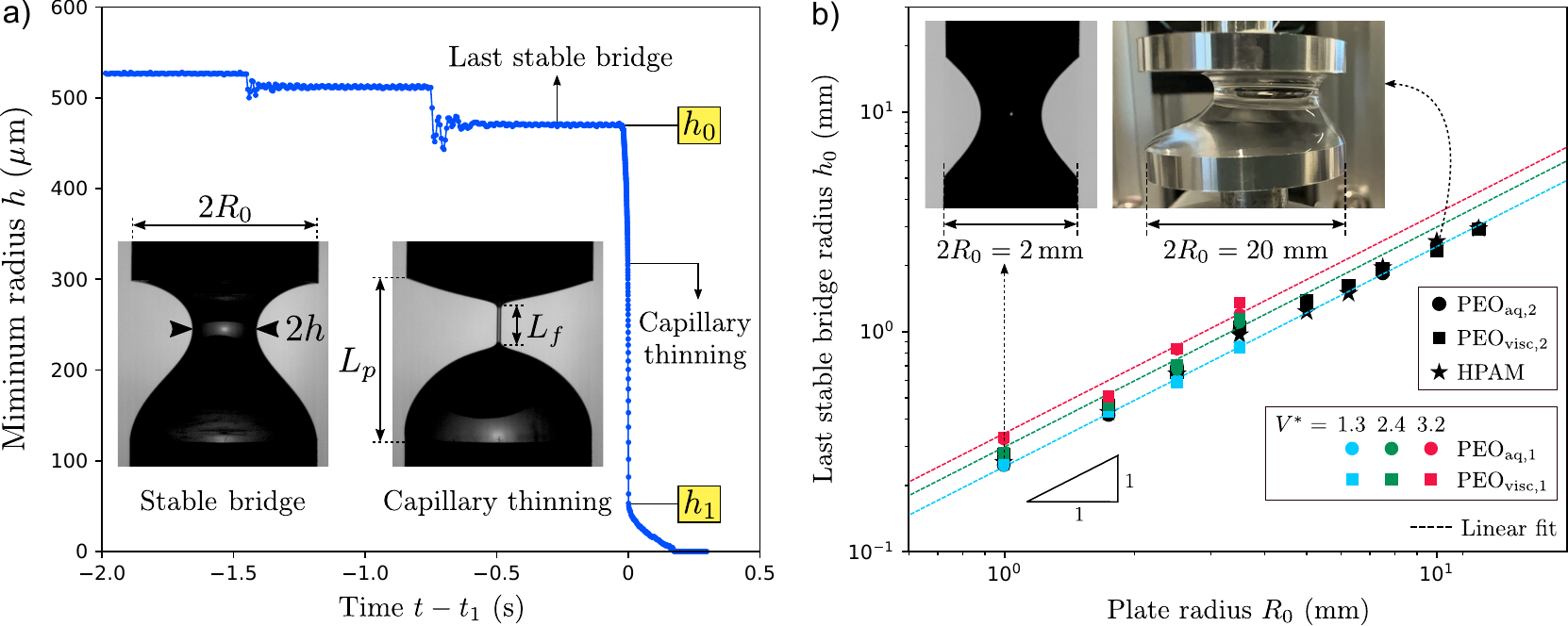}}
  \caption{(a) Time evolution of the minimum bridge / filament radius $h$ in our slow stepwise plate separation protocol for the PEO$_{\mathrm{aq,1}}$ solution for plate diameter $2R_0 = 3.5$~mm and a droplet volume $V^* = V/R_0^3 \approx 2.4$. Inertio-capillary oscillations are visible after each step. Inset images correspond to a stable liquid bridge (left) and to a thinning filament (right) of the same liquid with $2R_0 = 7$~mm and $V^* \approx 2.4$. (b) Last stable bridge radius $h_0$ against the plate radius $R_0$ for $2R_0$ between $2$ and $7$~mm and, for each plate, $V^* \approx 1.3$, $2.4$ and $3.2$ for the PEO$_{\mathrm{aq,1}}$ and PEO$_{\mathrm{visc,1}}$ solutions, and for $2R_0$ between $2$ and $25$~mm and an single volume ($V^* \approx 2.4$ for the smallest plates and $0.88$ for the largest plates) for the PEO$_{\mathrm{aq,2}}$ and PEO$_{\mathrm{visc,2}}$ and HPAM solutions. Inset images correspond to stable liquid bridges ($h \ge h_0$) for $2R_0 = 2$~mm (left, PEO$_{\mathrm{aq,1}}$ solution with $V^* \approx 2.4$) and $2R_0 = 20$~mm (right, HPAM solution with $V^* \approx 1.0$).}
\label{fig:step}
\end{figure}

The CaBER setup and slow-stepwise plate separation protocol described here are the same as in our previous paper \cite{gaillard2023beware}. A droplet of volume $V$ is placed on a horizontal plate of radius $R_0$ and the motor-controlled top plate of same radius is first moved down until it is fully wetted by the liquid, i.e., until the liquid bridge between the plates has a quasi cylindrical shape. The top plate is then moved up slowly (at about $0.5$~mm/s) and stopped at a plate separation distance $L_p$ where the liquid bridge is still stable, like in the left inset image of figure \ref{fig:step}(a), but close to the bridge instability threshold. Then, instead of moving the top plate at a constant (lower) velocity, i.e. like in SRM \cite{campo2010slow}, we move it by 10~$\mu$m $L_p$-increment steps, waiting about one second between each step (longer than the solution's relaxation time), which is long enough to ensure that polymers are at equilibrium (no pre-stress) before each new step. At a certain step, the bridge becomes unstable and collapses under the action of surface tension, transiently leading to the formation of a nearly cylindrical filament which is the signature of viscoelastic pinch-off, as shown in the right inset image of figure \ref{fig:step}(a). We stop moving the top plate once capillary-driven thinning starts. 

The process is recorded by a high-magnification objective mounted on a high-speed camera (Phantom TMX 7510) and images are analysed by a Python code. A typical time-evolution of the minimum bridge / filament radius, labelled `$h_{\mathrm{min}}$' by many authors but simply referred to as $h$ in the rest of the paper, is shown in figure \ref{fig:step}(a). The purpose of this step-by-step plate separation protocol is to extract the value of the last stable bridge radius $h_0$ which, since steps are small, can be considered as the initial bridge radius at the onset of capillary thinning. Our image resolution is up to 1 pixel per micrometer for the smallest drops, corresponding to the smallest plates, and our time resolution is $15,000$ images per second to capture the fast bridge collapse from radius $h_0$ to the radius $h_1$ marking the onset of the elastic regime, see figure \ref{fig:step}(a).

The critical aspect ratio $\Lambda = L_p/(2 R_0)$ at which the liquid bridge becomes unstable depends on the liquid volume $V$ and on the Bond number $\Bo = \rho g R_0^2 / \gamma$, where $g$ is the gravitational acceleration \cite{slobozhanin1993stability}. In our experiments, we vary the plate diameter $2R_0$, between $2$ and $25$~mm, as well as the non-dimensional droplet volume 
\begin{equation}
V^* = V/R_0^3.
\label{eq:V*}
\end{equation}
\noindent As shown in figure \ref{fig:step}(b), the last stable bridge radius increases fairly linearly with the plate radius, i.e. $h_0 \propto R_0$ with a prefactor which increases with $V^*$, with no strong dependence on the liquid used since they all have comparable surface tensions. Typically, $h_0/R_0$ ranges between $0.24$ and $0.35$ for $V^* \approx 1.3$ and $3.2$, respectively. 

The aluminium plates are plasma-treated before each measurement to increase their hydrophilicity and hence prevent dewetting of the top plate. However, dewetting could not be avoided for plate diameters $2R_0 \ge 10$~mm, as shown by the right inset image of figure \ref{fig:step}(b) showing a stable liquid bridge ($h \ge h_0$) where the top end-drop does not cover the top plate fully for $2R_0 = 20$~mm. Perhaps surprisingly, $h_0$ does not saturate at $2R_0 \ge 10$~mm in spite of this lack of full coverage, see figure \ref{fig:step}(b). For such large plates, the top end-drop is not necessarily at the centre of the the top plate since the two plates are not perfectly parallel. Note that because of the plasma treatment, there is always a thin film covering the top plate. 

All experiments are carried out at a high relative humidity ($> 80$\%) ensured by placing the CaBER setup in a box with wet paper tissues. We checked that repeating an experiment several times over the course of $10$ minutes does not lead to any monotonic increase or decrease of the filament thinning rate ($1/3\tau_e$, see \S\ref{sec:Experimental results on $h_1$}) over time in the exponential part of the elastic regime, beyond small variations of less than $5$\%, suggesting that both evaporation and polymer degradation (which may occur during bridge / filament thinning) are negligible.

\subsection{Equations and numerical methods}
\label{subsec:Numerical methods}

The numerical simulations discussed in \S\ref{sec:Oldroyd-B prediction for $h_1$} and \S\ref{sec:FENE-P prediction for $h_1$} are performed using the FENE-P model which aims to describe polymer stretching as well as the finite extensibility of polymer molecules. We consider a cylindrical axisymmetric $(r,z)$ coordinate system aligned with the vertical axis of the liquid bridge. In the simulations, we integrate the mass and momentum conservation equations of general form
\begin{equation}
 \nabla \cdot {\bf v} = 0,
\end{equation}
\begin{equation}
\rho\frac{D{\bf v}}{Dt}  = -\nabla p+\nabla\cdot\boldsymbol \sigma,
\label{eq:miguel_Cauchy}
\end{equation}
\noindent where $\rho$, ${\bf v}=v_r({ r,z},t) {\bf e}_r+v_z(r,z,t) {\bf e}_z$, and $p({ r,z},t)$ are the density, velocity and (reduced) pressure field (accounting for gravity), respectively, and $D/Dt$ is the material derivative. These equations are completed with the constitutive relationships for the stress tensor $\boldsymbol \sigma = \boldsymbol \sigma_s+\boldsymbol \sigma_p$, where 
\begin{equation}
\boldsymbol \sigma_s = \eta_s \left(\nabla{\bf v} + (\nabla{\bf v} )^\mathrm{T}\right)
\end{equation}
\noindent is the contribution of the solvent of viscosity $\eta_s$ and where $\boldsymbol \sigma_p$ is the polymer contribution. In the FENE-P model \cite{SPHE20}, this contribution is calculated as
\begin{equation}
{\boldsymbol \sigma}_p = \frac{\eta_p}{\tau} (f{\bf A}-{\bf I}), 
\quad f = \frac{1}{1-\text{tr}({\bf A})/L^2},
\label{eq:miguel_sigmap}
\end{equation}
\noindent where $\eta_p$ is the polymer contribution to the zero-shear viscosity $\eta_0 = \eta_s + \eta_p$, $\tau$ is the relaxation time and $L^2$ is the finite extensibility limit, ${\bf I}$ being the identity matrix. The conformation tensor $\mathbf{A}$ is calculated from the nonlinear relaxation law
\begin{equation}
%{\stackrel{\triangledown}{\mathbf A}} = - \frac{1}{\tau}(f {\bf A}-{\bf I}),
\frac{\mathrm{D} {\bf A}}{\mathrm{D} t} - \left[ (\nabla{\bf v})^\mathrm{T} \cdot {\bf A} + {\bf A} \cdot \nabla{\bf v} \right]
= - \frac{1}{\tau}(f {\bf A}-{\bf I}).
\label{eq:miguel_FENEP}
\end{equation} 

% Interface boundary conditions
The free surface location is defined by the equation $r=h(z,t)$. The boundary conditions at that surface are:
\begin{equation}
\frac{\partial h}{\partial t}+h_z w-u=0,
\label{int1}
\end{equation}
\begin{eqnarray}
-p+gz-\frac{hh_{zz}-1-h_z^{2}}{h(1+h_z^{2})^{3/2}}+{\bf n}\cdot {\bf \sigma}\cdot {\bf n}=0,
\label{eq:int3}
\end{eqnarray}
\begin{equation}
{\bf t}\cdot {\bf \sigma}\cdot {\bf n}=0, 
\label{eq:int2}
\end{equation}
where $h_z \equiv dh/dz$, $h_{zz} \equiv  d^2h/dz^2$, $g$ is the gravitational acceleration, ${\bf n}$ is the unit outward normal vector and ${\bf t}$ is the unit vector tangential to the free surface meridians. Equation (\ref{int1}) is the kinematic compatibility condition, while equations (\ref{eq:int3}) and (\ref{eq:int2}) express the balance of normal and tangential stresses  respectively. The anchorage condition $h=R_0$ is set at $z=0$ and $z=L_p$ where $L_p$ is the plate separation distance. The nonslip boundary condition is imposed at the solid surfaces in contact with the liquid. The volume $V$ of the initial configuration is prescribed (and conserved), namely,
\begin{equation}
\pi\int_{0}^{L} h^2\ dz=V.\label{volume}
\end{equation}

We start the simulation from liquid bridge at equilibrium with a plate separation distance $L_p$ just below (very close to) the critical one. The breakup process is triggered by applying a very small gravitational force perturbation. We note $h_0$ the minimum radius of the (stable) liquid bridge just before the perturbation is applied, which is conceptually identical to the last stable bridge radius described in \S\ref{subsec:CaBER setup and protocol} for a slow stepwise plate separation protocol.

A numerical simulation is fully determined by five quantities: the Ohnesorge number, the Deborah number, the non-dimensional droplet volume $V^* = V/R_0^3$, the finite extensibility parameter $L^2$ and the viscosity ratio 
\begin{equation}
S = \eta_s / \eta_0.
\label{eq:S}
\end{equation}
\noindent While equations are non-dimensionalised using $R_0$ and $\sqrt{\rho R_0^3 / \gamma}$ as the characteristic length and time scales, we only refer in \S\ref{sec:Oldroyd-B prediction for $h_1$} and \S\ref{sec:FENE-P prediction for $h_1$} to values involving $h_0$ since, as we will show in \S\ref{subsec:Influence of the plate radius and drop volume}, it is $h_0$ (not $R_0$) which is the most relevant length scale of the problem. In particular, we refer to the Rayleigh and viscous time scales and Ohnesorge and Deborah numbers defined in equations \ref{eq:tau_R} to \ref{eq:De}.

Simulations are preformed in the absence of gravity where the threshold of the Rayleigh-Plateau instability, and therefore the shape of the initial bridge of minimum radius $h_0$ from which simulations start, is solely determined by $V^*$. In \S\ref{sec:Oldroyd-B prediction for $h_1$} and \S\ref{sec:FENE-P prediction for $h_1$}, instead of referring to $V^*$, we refer to the value of $h_0/R_0$ since, in experiments, $h_0/R_0$ is not only set by $V^*$ but also by the Bond number. We recall that, in experiments, $h_0/R_0$ ranges between $0.24$ and $0.35$ for $V^* \approx 1.3$ and $3.2$, respectively, see figure \ref{fig:step}(b). 
%and we expect simulations to potentially match experiments only for matching values of $h_0/R_0$, not for matching values of $V^*$. 

The model was solved with a variation of the method described in \cite{herrada2016numerical}. The physical domain occupied by the liquid is mapped onto a rectangular domain through a coordinate transformation. Each variable and its spatial and temporal derivatives appearing in the transformed equations were written as a single symbolic vector. Then, we used a symbolic toolbox to calculate the analytical Jacobians of all the equations with respect to the symbolic vector. Using these analytical Jacobians, we generated functions that could be evaluated in the iterations at each point of the discretised numerical domains. 

The transformed spatial domain is discretised using $n_\eta=11$ Chebyshev spectral collocation points in the transformed radial direction $\eta$ of the  domain. We used $n_\xi=501$ equally spaced collocation points in the transformed axial direction $\xi$. The axial direction was discretised using second finite differences. Second-order backward finite differences were used to discretise the time domain. We used an automatic variable time step based on the norm of the difference between the solution calculated with a first-order approximation and that obtained from the second-order procedure. The nonlinear system of discretised equations was solved at each time step using the Newton method. The method is fully implicit.

%%%%%%%%%%%%%%%%%%%%%%%%%%%%%%%%%%%%%%%%%%%%%%%%%%%%%%%%%%%%%%%%%%%%%%%%%%%%%%%%%%%%%%%%%%%%%%%%%%%%%%%%%%%%%%%%%%%%%%%%%%%%%%%%
\section{Experimental results}
\label{sec:Experimental results on $h_1$}

In this experimental section, we investigate the role of the plate radius and sample volume in \S\ref{subsec:Influence of the plate radius and drop volume} and of the polymer concentration in \S\ref{subsec:Influence of the polymer concentration} on the pinch-off dynamics.

\subsection{Influence of the plate radius and drop volume}
\label{subsec:Influence of the plate radius and drop volume}

\begin{figure}
  \centerline{\includegraphics[scale=0.58]{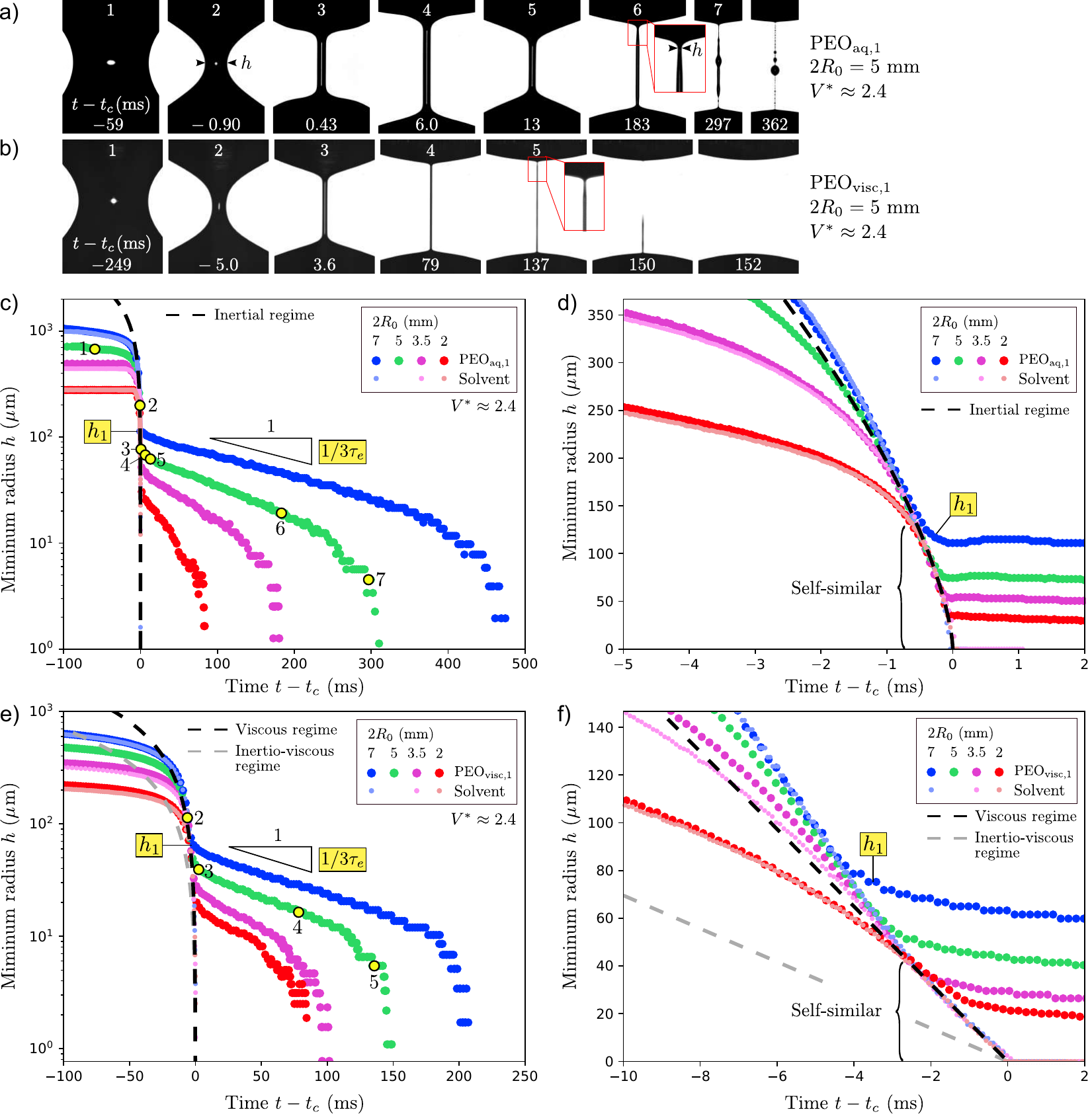}}
  \caption{(a,b) Image sequences of the bridge / filament for the PEO$_{\mathrm{aq,1}}$ (a) and PEO$_{\mathrm{visc,1}}$ (b) solutions tested with plate diameter $2R_0 = 5$~mm and droplet volume $V^* \approx 2.4$. (c-f) Time evolution of the minimum bridge / filament radius $h$ in semi-log (c,e) and lin-lin (d,f), for plate diameters $2R_0$ between $2$ and $7$~mm and fixed $V^* \approx 2.4$, for the PEO$_{\mathrm{aq,1}}$ (c,d) and PEO$_{\mathrm{visc,1}}$ (e,f) solutions and for their respective solvents (smaller data points), compared with equations \ref{eq:selffimilar_inviscid} and \ref{eq:selffimilar_viscous} where $t_c$ is the solvent breakup time. Times with labels 1-7 and 2-5 in panels (c) and (e) for $2R_0=5$~mm correspond to the snapshots in panels (a) and (b).}
\label{fig:ht_water_and_PEG30}
\end{figure}

Image sequences of the pinch-off dynamics are shown in figures \ref{fig:ht_water_and_PEG30}(a) and \ref{fig:ht_water_and_PEG30}(b) for the PEO$_{\mathrm{aq,1}}$ ($500$~ppm PEO-4M in water) solution and the PEO$_{\mathrm{visc,1}}$ ($25$~ppm PEO-4M in a $\sim 260$ more viscous solvent) solution, respectively, illustrating the transition from a bridge shape in the Newtonian regime to a filament shape in the elastic regime. For the PEO$_{\mathrm{aq,1}}$ solution, the filament is initially cylindrical until localised pinching is observed near one of the of the end drops (see frame 6 in figure \ref{fig:ht_water_and_PEG30}(a)), followed by its destabilisation into a succession of beads connected by thin filaments (which are below our spatial resolution), a phenomenon usually referred to as ``blistering'' instability \cite{sattler2008blistering,sattler2012final,eggers2014instability,semakov2015nature} (see frame 7 and 8). For the PEO$_{\mathrm{visc,1}}$ solution, local pinching occurs very close to breakup and no blistering is observed. In this paper, we refer to the minimum bridge / filament radius $h$ which therefore corresponds to the pinched region if localised pinching occurs. On another note, inertio-capillary oscillations of the top and bottom end drop lead to oscillations of the filament length for the PEO$_{\mathrm{aq,1}}$ solution (see frames 3 to 6 in figure \ref{fig:ht_water_and_PEG30}(a)). These oscillations are absent for the PEO$_{\mathrm{visc,1}}$ solution due to viscous damping. Note that oscillations do not lead to significant oscillations of the filament radius, implying that when the length of the filament increases, new filament is being created from the liquid in the end drops.

Time evolutions of the minimum bridge / filament radius $h$ are shown in figures \ref{fig:ht_water_and_PEG30}(c-f) for four plate diameters between $2$ and $7$~mm and a fixed non-dimensional drop volume $V^* \approx 2.4$ for the PEO$_{\mathrm{aq,1}}$ solution (figures \ref{fig:ht_water_and_PEG30}(c) and \ref{fig:ht_water_and_PEG30}(d)) and the PEO$_{\mathrm{visc,1}}$ solution (figures \ref{fig:ht_water_and_PEG30}(e) and \ref{fig:ht_water_and_PEG30}(f)) in semi-log (figures \ref{fig:ht_water_and_PEG30}(c) and \ref{fig:ht_water_and_PEG30}(e)) and lin-lin (figures \ref{fig:ht_water_and_PEG30}(d) and \ref{fig:ht_water_and_PEG30}(f)), the latter focusing on the transition to the elastic regime. The smaller data points correspond to the solvent alone for three of the same plate diameters and, in each case, the same $V^*$ (to within experimental reproducibility). The time reference $t_c$ corresponds to the critical time at which the bridge of solvent alone breaks up. For polymer solutions, since $t_c$ can not be determined, curves are shifted along the time axis until overlapping their corresponding solvent curves. The good overlap between polymer solutions and their solvent at all times $t<t_1$ (before the transition to the elastic regime) confirms that polymers do not affect the pinch-off dynamics in the (hence rightfully called) Newtonian regime. For the PEO$_{\mathrm{visc,1}}$ solution where, as is about to be discussed, capillarity is balanced by viscosity in the Newtonian regime, this solution-solvent overlap is consistent with the low polymer contribution to the total shear viscosity ($\eta_p / \eta_0 = 0.013$). The least good solution-solvent overlaps are explained by experimental differences in $V^*$. 

All curves corresponding to Newtonian solvents in figure \ref{fig:ht_water_and_PEG30}(d,f) overlap close to breakup, indicating a self-similar thinning regime where the information about initial condition, set by $R_0$ and $V^*$, is forgotten. Such overlap is also observed for droplets of different volumes for a given plate radius. For the water solvent in figure \ref{fig:ht_water_and_PEG30}(d), the self-similar regime is well captured by the inertio-capillary thinning law
\begin{equation}
h = A \left( \frac{\gamma}{\rho} \right)^{1/3} (t_c-t)^{2/3},
\label{eq:selffimilar_inviscid}
\end{equation}
\noindent with a prefactor $A = 0.47$ which is consistent with the experimental and numerical results of Deblais et al. \cite{deblais2018viscous}. For the $\sim 260$ times more viscous solvent in figure \ref{fig:ht_water_and_PEG30}(f), the self-similar regime is well captured by the visco-capillary thinning law \cite{papageorgiou1995breakup,mckinley2000extract}
\begin{equation}
h = 0.0709 \frac{\gamma}{\eta_0} (t_c-t).
\label{eq:selffimilar_viscous}
\end{equation}
\noindent This is consistent with the fact that the Ohnesorge number $\Oh = \eta_0/\sqrt{\rho \gamma h_0}$ (see equation \ref{eq:Oh}) is up to $0.02$ for PEO$_{\mathrm{aq}}$ and up to $0.1$ for HPAM (for the smallest plate diameter where $h_0$ is lowest), i.e. $\Oh \ll 1$, and ranges between $1.0$ and $1.9$ for the PEO$_{\mathrm{visc,1}}$ solution in the range of plate diameters considered in figure \ref{fig:ht_water_and_PEG30}. In spite of the moderate Ohnesorge numbers in the latter case, we do not observe a clear transition to the inertio-visco-capillary thinning law $h = 0.0304 (\gamma/\eta_0) (t_c-t)$ \cite{eggers1993universal,eggers1997nonlinear} which describes the behaviour of Newtonian fluids close to breakup, see figure \ref{fig:ht_water_and_PEG30}(d).

Interestingly, the transition to the elastic regime occurs around the time at which the self-similar Newtonian regime is reached in figure \ref{fig:ht_water_and_PEG30}(d,f), slightly after for the PEO$_{\mathrm{aq,1}}$ solution and slightly before for the PEO$_{\mathrm{visc,1}}$ solution. However, in both cases, the transition radius $h_1 = h(t_1)$ increases with the plate diameter $2R_0$, which indicates that polymers already started to  significantly deform before the self-similar regime. Indeed, if polymers only started to deform within the self-similar regime where the thinning dynamics does not depend on $R_0$ or $V^*$ anymore, the amount of polymer deformation would be independent on the initial condition, leading to a transition radius $h_1$ which would not depend on $R_0$ or $V^*$, as we discuss further in \S\ref{subsec:Oldroyd-B Simulations}.

%The end of the exponential regime occurs at a time $t_2$ corresponding to at a filament radius $h_2 = h(t_2)$ which also increases with $R_0$, see figure \ref{fig:ht_water_and_PEG30}(a,c). We define $t_2$ as the time where $h$ differs by more than 15\% from an extrapolation of the exponential regime. To improve the estimation, the extrapolation is in fact compared with an empirical fit of $h(t)$ for $t>t_1$ of the form $h(t) = a_1 \exp{(-b_1t)} -c_1t + d_1$ proposed by Anna \& McKinley \cite{anna2001elasto}, where $a_1$, $b_1$, $c_1$ and $d_1$ are fitting parameters. This is illustrated in figure \ref{fig:ht_water_and_PEG30}(a) for $2R_0 = 7$~mm. 

After the filament formation, the thinning rate $\vert \dot{h}/h \vert$ is initially fairly constant, indicating an exponential decay, and increases close to breakup in a so-called `terminal regime' where authors argue that polymer chains approach full extension and a Newtonian-like high-viscosity dynamics is recovered \cite{dinic2019macromolecular,anna2001elasto,campo2010slow,stelter2002investigation}. The (constant) filament thinning rate measured during the exponential part of the elastic regime is found to decrease with increasing plate diameter, see figure \ref{fig:ht_water_and_PEG30}(c,e). This is inconsistent with the Oldroyd-B model which predicts $\vert \dot{h}/h \vert = 1/3\tau$ (see equation \ref{eq:exponential}) where $\tau$ is the (longest) relaxation time of the polymer solution which is a fluid property, independent of the size of the system. As we show in our previous paper \cite{gaillard2023beware}, this surprising dependence on the system size is also observed for the classical step-strain plate separation protocol of a commercial CaBER rheometer as well as for Dripping-onto-Substrate and dripping \cite{rajesh2022transition} experiments. We show that this is not caused by artefacts such as solvent evaporation or polymer degradation, suggesting that the liquid does not change when being tested with different plate diameters. To discuss this geometry-dependent filament thinning rate, we define an apparent (or effective) relaxation time $\tau_e$ such that $\vert \dot{h}/h \vert = 1/3\tau_e$ during the exponential part of the elastic regime.

\begin{figure}
  \centerline{\includegraphics[scale=0.58]{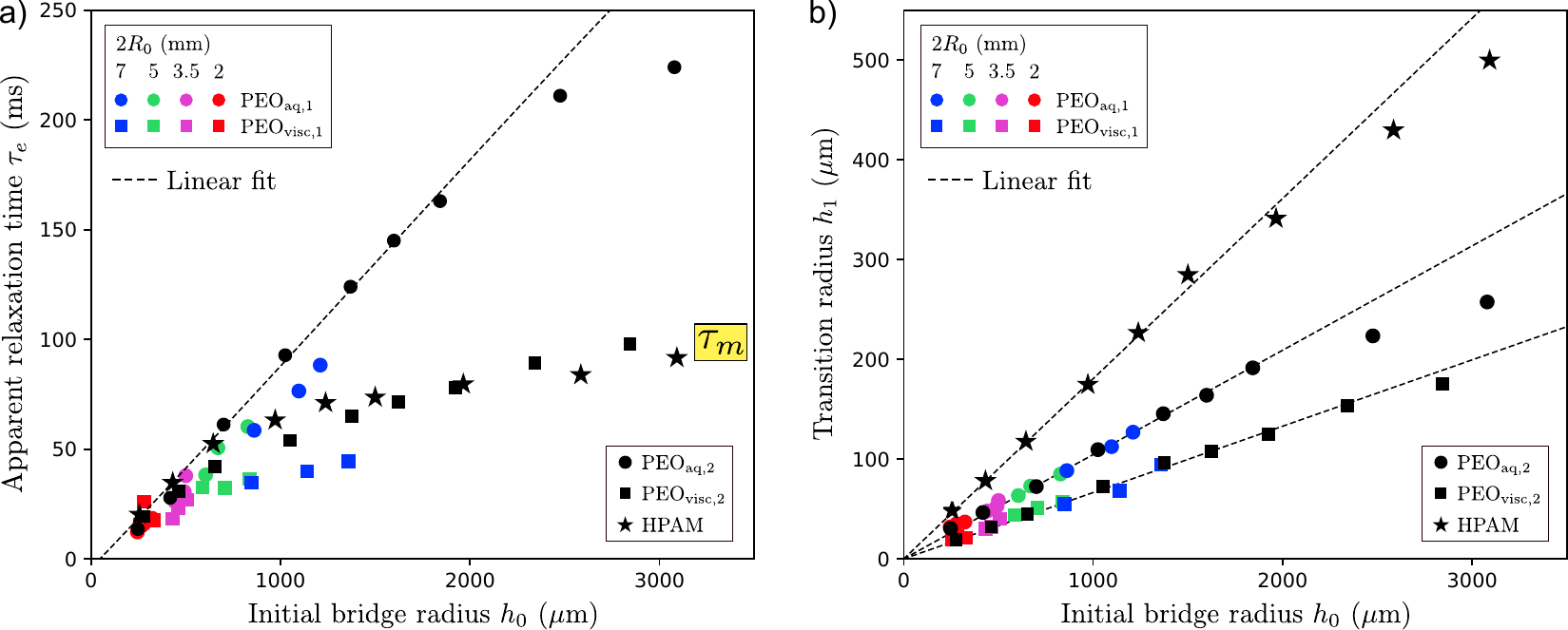}}
  \caption{Effective extensional relaxation time $\tau_e$ (a) and transition radius $h_1$ (b) against the last stable bridge radius $h_0$ for different plate radii $R_0$ and droplet volumes $V^*$ for all polymer solutions. For the PEO$_{\mathrm{aq,1}}$ and the PEO$_{\mathrm{visc,1}}$ solutions, three points of the same colour correspond to the same $R_0$ and three different $V^*$.}
\label{fig:h12_taue}
\end{figure}

The apparent relaxation time $\tau_e$ and the transition radius $h_1$ are plotted against $h_0$ in figure \ref{fig:h12_taue} for different polymer solutions, plate diameters $2R_0$ and droplet volumes $V^*$. The fact that data corresponding to different values of $R_0$ and $V^*$ collapse on a single curve for both the PEO$_{\mathrm{aq,1}}$ and the PEO$_{\mathrm{visc,1}}$ solutions suggests that $h_0$, which is an increasing function of both $R_0$ and $V^*$ (see figure \ref{fig:step}(b)) is the only relevant geometrical parameter of the problem. This is the reason why we chose $h_0$ at the relevant length scale for non-dimensional numbers such as the Ohnesorge and Deborah numbers in equations \ref{eq:Oh} and \ref{eq:De}. This is in agreement with the idea that the thinning dynamics is only influenced by extensional flow in the bridge / filament while the top and bottom end droplets act as passive liquid reservoirs. 

We find that $\tau_e$ seems to saturate towards a maximum value $\tau_m$ at large $h_0$, see figure \ref{fig:h12_taue}(a). The estimated values of $\tau_m$ are reported in table \ref{tab:rheology_bigpins} for the PEO$_{\mathrm{aq,2}}$, PEO$_{\mathrm{visc,2}}$ and HPAM solutions for which plate diameters up to $25$~mm were used, well beyond typical CaBER plate sizes, which was needed to observe the saturation of $\tau_e$. In our previous paper \cite{gaillard2023beware}, we explored the possibility that $\tau_m$ could be the `real' relaxation time of the solution, invoking finite extensibility effects described by the FENE-P model to explain thinning rates larger than $1/3\tau_m$ for low $h_0$. We concluded that this was a possible explanation only for the PEO$_{\mathrm{visc,2}}$ solution but not for the PEO$_{\mathrm{aq,2}}$ and HPAM solutions, suggesting that the FENE-P model misses some important features of polymer dynamics in extensional flows.

The first transition radius $h_1$ increases fairly linearly with $h_0$ for all liquids, see figure \ref{fig:h12_taue}(b). This is in contradiction with the scaling $h_1 \propto h_0^{4/3}$ expected from the Oldroyd-B model when assuming that polymer relaxation is negligible during the time needed for the bridge to thin from $h_0$ to $h_1$, see equation \ref{eq:h1_norelaxation}. Since $h_0 \propto R_0$ for a fixed $V^*$ (see figure \ref{fig:step}(b)), this implies that $h_1 \propto R_0$, different from the scaling $h_1 \propto R_{n}^{0.66}$ observed experimentally by Rajesh et al. \cite{rajesh2022transition} in the analogous problem of a drop falling from a nozzle of radius $R_n$. This is surprising since $R_n$ should play the same role as the plate radius $R_0$ in CaBER.

%The second transition radius $h_2$ also increases with $h_0$, typically as $h_2 \propto h_0^{k_2}$ where $k_2 <1$. More precisely,  $k_2 \approx 0.89$, $0.56$ and $0.65$ for the PEO$_{\mathrm{aq,2}}$, PEO$_{\mathrm{visc,2}}$ and HPAM solutions, respectively. This implies that $h_1/h_2 \propto h_0^{1-k_2}$ which means that the exponential part of the elastic regime, where $\tau_e$ is measured, covers an increasingly large range of filament radii as $h_0$ increases, see figure \ref{fig:h12_taue}(b). 

Note that the PEO$_{\mathrm{aq,2}}$ and PEO$_{\mathrm{visc,2}}$ solutions are slightly more elastic than the PEO$_{\mathrm{aq,1}}$ and PEO$_{\mathrm{visc,1}}$ solutions since they exhibit larger apparent relaxation times, see figure \ref{fig:h12_taue}(a). However, these differences are barely visible in \ref{fig:h12_taue}(b) since the values of $h_1$ are almost the same, suggesting that the dependence of $h_1$ on $\tau_e$ is relatively weak, as we now confirm by varying the polymer concentration.

\subsection{Influence of the polymer concentration}
\label{subsec:Influence of the polymer concentration}

\begin{figure}
  \centerline{\includegraphics[scale=0.58]{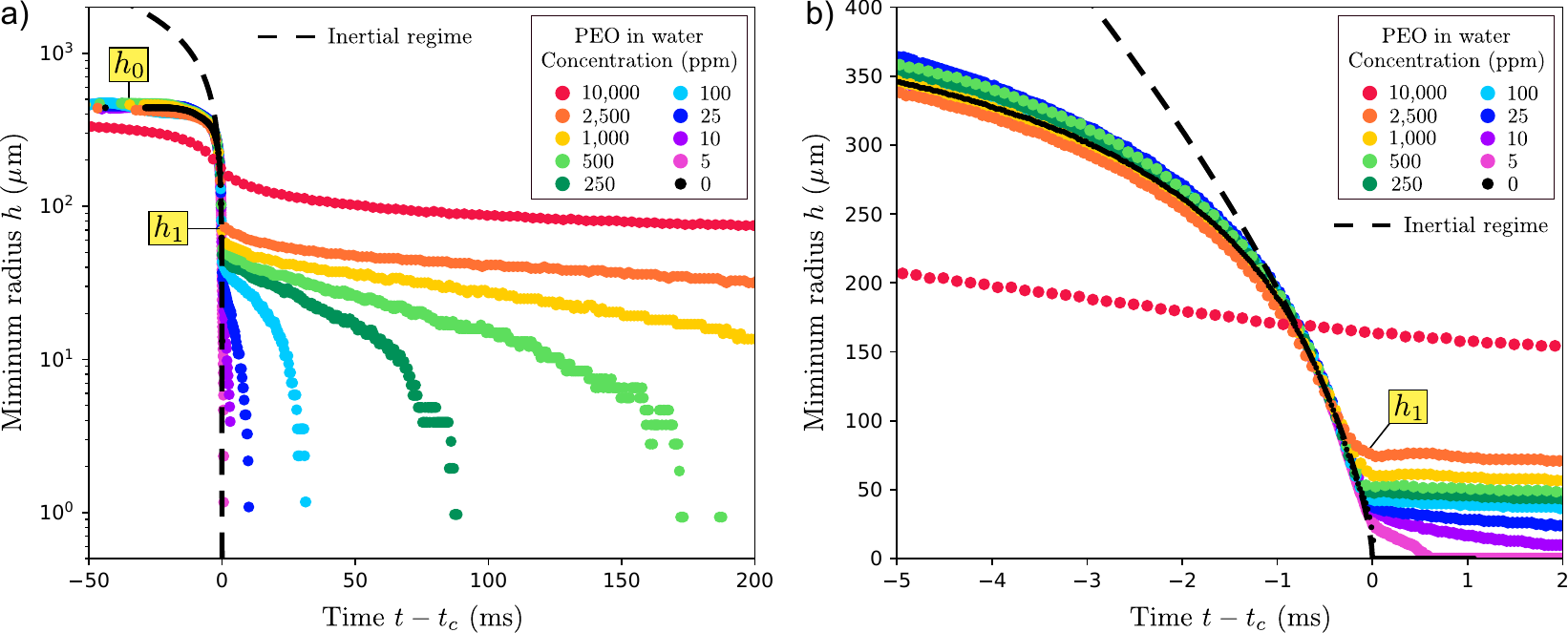}}
  \caption{Time evolution of the minimum bridge/filament radius $h$ in lin-log (a) and lin-lin, focusing on the first transition (b), for PEO-4M solutions of different concentrations in water, for a fixed plate diameter $2R_0 = 3.5$~mm and drop volume $V^* \approx 2.4$. The time $t_c$ is the time at which the bridge would break up for the solvent alone, here water.}
\label{fig:ht_varyc}
\end{figure}

Figure \ref{fig:ht_varyc} shows the time evolution of the minimum bridge / filament radius for the aqueous PEO solutions of table \ref{tab:rheology_PEO} of various PEO concentrations, water solvent included, which were tested for a fixed plate diameter $2R_0 = 3.5$~mm and droplet volume $V^* \approx 2.4$, in semi-log (figure \ref{fig:ht_varyc}(a)) and in lin-lin focusing on the transition to the elastic regime (figure \ref{fig:ht_varyc}(b)). Like figure \ref{fig:ht_water_and_PEG30}, the time $t_c$ at which the solvent breaks is chosen as the time reference and curves corresponding to polymer solutions are shifted along the time axis to maximise the overlap with the solvent for $t<t_1$. This overlap is very good up to $2,500$~ppm (dilute and unentangled semi-dilute solution), small deviations being attributable to slightly different droplet volumes. For $10,000$~ppm however (entangled semi-dilute solutions), the bridge dynamics prior to the exponential regime is radically different from the pure solvent case, indicating that elasticity is not negligible even before the exponential regime is reached.

\begin{figure}[t]
  \centerline{\includegraphics[scale=0.58]{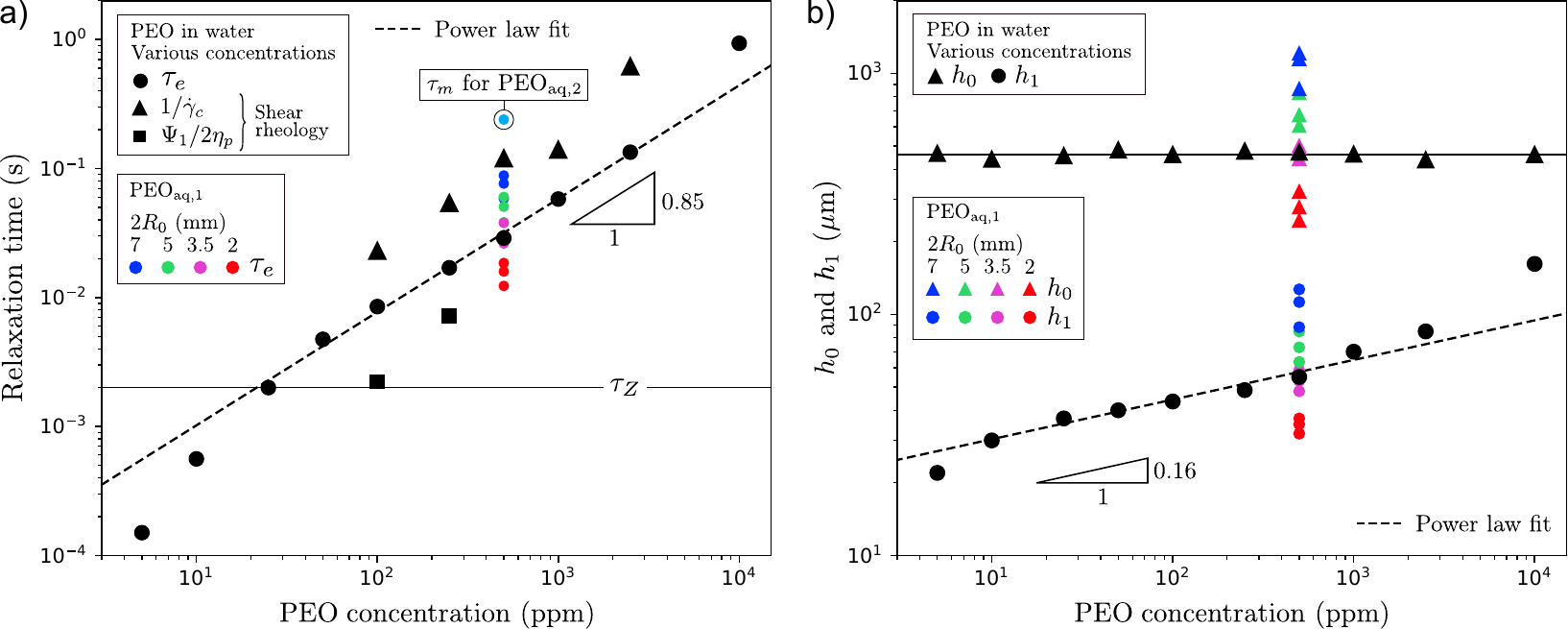}}
  \caption{(a) Relaxation times and (b) transition radius $h_1$ and last stable bridge radius $h_0$ against polymer concentration for PEO-4M solutions of different concentrations in water. The data of panel (b) and the effective CaBER relaxation time $\tau_e$ in panel (a) correspond to a fixed plate diameter $2R_0 = 3.5$~mm and drop volume $V^* \approx 2.4$, except for the 500~ppm solution where data corresponding to different $2R_0$ (between $2$ and $7$~mm) and $V^*$ are shown. In panel (a), we also plot the relaxation time $1/\dot{\gamma}_c$ and $\Psi_1/ 2 \eta_p$ from shear rheology, as well as the Zimm relaxation time $\tau_Z$. We also show the maximum CaBER relaxation time $\tau_m$ (high $h_0$ limit of $\tau_e$) measured for the ($500$~ppm) PEO$_{\mathrm{aq,2}}$ solution (which is slightly more elastic than the PEO$_{\mathrm{aq,1}}$ solution from which the other $500$~ppm data points are from, see figure \ref{fig:h12_taue}(a)).}
\label{fig:h012_tau_varyc}
\end{figure}

The apparent relaxation time $\tau_e$ and the transition radius $h_1$ are plotted in figure \ref{fig:h012_tau_varyc} as a function of the polymer concentration $c$ for the aqueous PEO solutions of table \ref{tab:rheology_PEO}. Since these solutions were tested for a fixed plate diameter $2R_0 = 3.5$~mm and droplet volume $V^* \approx 2.4$, the initial bridge radius $h_0 \approx 460$~$\mu$m, also plotted in figure \ref{fig:h012_tau_varyc}(b), is the same for each solution. For the $500$~ppm solution, labelled PEO$_{\mathrm{aq,1}}$, which was tested for four plate diameters from $2$ to $7$~mm and, in each case, with three different volumes, the corresponding values of $\tau_e$, $h_0$ and $h_1$ are also included in figure \ref{fig:h012_tau_varyc} to illustrate how data for a given concentration can shift as $h_0$ is varied. Although not large enough plate diameters were used to estimate the high-$h_0$ limit value $\tau_m$ of the relaxation time for the PEO$_{\mathrm{aq,1}}$ solution, the value estimated for the slightly more elastic PEO$_{\mathrm{aq,2}}$ solution (see figure \ref{fig:h12_taue}(a) and table \ref{tab:rheology_bigpins}) is shown in \ref{fig:h012_tau_varyc}(a) for $500$~ppm.

We find a weak dependence $h_1 \propto c^{0.16}$ for dilute and semi-dilute solutions, see figure \ref{fig:h012_tau_varyc}(b), consistent with Rajesh et al. \cite{rajesh2022transition} who found $h_1 \propto c^{0.15}$ in the analogous problem of a drop falling from a nozzle. We find a power law $\tau_e \propto c^{0.85}$ for the apparent relaxation time in figure \ref{fig:h012_tau_varyc}(a), consistent with the wide range of exponents, typically between $0.6$ and $1$, reported by other authors in CaBER \cite{bazilevsky1997failure,zell2010there,clasen2006dilute} and pendant drop experiments \cite{tirtaatmadja2006drop,rajesh2022transition}. Note that this exponent is expected to be a function of the solvent quality \cite{clasen2006dilute} and that its potential dependence on $h_0$ is not investigated in the present work. From these scalings, we deduce a weak dependence of $h_1 \propto \tau_e^{0.19}$ on the apparent relaxation time.

The apparent CaBER relaxation time $\tau_e$ in figure \ref{fig:h012_tau_varyc}(a) is compared to values estimated from shear rheology, i.e. $1/\dot{\gamma}_c$ which is estimated from shear viscosity curves featuring shear-thinning, which excludes low concentrations, and $\Psi_1 / 2 \eta_p$ which is estimated from the first normal stress difference for the only two measurable solutions exhibiting a quadratic scaling $N_1 \propto \dot{\gamma}^2$. We observe that $1/\dot{\gamma}_c$ follows a power law with an exponent close to the one found for $\tau_e$. Additionally, we find that $\tau_e$ is larger than $\Psi_1 / 2 \eta_p$ in figure \ref{fig:h012_tau_varyc}(a) for this specific plate radius, and would hence be even larger in the high-$h_0$ limit $\tau_m$. Hence, even for dilute solutions exhibiting weak shear-thinning and quadratic normal stresses, for which the Oldroyd-B model could describe the shear rheology, there is no quantitative agreement between the relaxation time measured from normal stresses and from filament thinning rheometry, in agreement with Zell et al. \cite{zell2010there} who dedicated a paper specifically on the link between $\tau_e$ and $\Psi_1$. The discrepancy between relaxation times measured in shear and in extension likely (partially) stems from the fact that elastic dumbbell models such as Oldroyd-B and FENE-P predict that polymer chains approach full extension in strong shear flows ($\tau \dot{\gamma} \gg 1$), in contradiction with the tumbling-motion-induced partial extension observed experimentally \cite{smith1999single}. In a perspective paper currently under preparation, Boyko \& Stone suggest that comparing shear and extension-derived model parameters should therefore be done using more complex models such as the FENE-PTML model (so named by the authors) proposed by Phan-Thien, Manero, and Leal \cite{phan1984study} which captures this partial extension under shear. In this paper, we only consider filament thinning extensional flows for which Oldroyd-B and FENE-P are suitable model candidates.

%The link between $\dot{\gamma}_c$ and the relaxation time of a polymer solution is however not obvious. In the FENE-P model, where shear-thinning arises from the finite extensibility of non-interacting (dilute) polymer chains, shear thinning starts at $\dot{\gamma}_c \approx L/\tau$ where $\tau$ is the relaxation time and $L$ is the ratio of the size of the fully extended chain to its equilibrium size \cite{gaillard2018flow}. We hence expect that $1/\dot{\gamma}_c \approx \tau / L \ll \tau$ for long chains for which $L \gg 1$, in disagreement with the data of figure \ref{fig:h012_tau_varyc}(a) where $1/\dot{\gamma}_c > \tau_e$. Assuming that the `real' relaxation time is in fact the high-$h_0$ limit $\tau_m$ (see figure \ref{fig:h12_taue}(a)), and taking $\tau_m \approx 240$~ms for the PEO$_{\mathrm{aq,1}}$ solution, the same as for the PEO$_{\mathrm{aq,2}}$ solution, we get $L \approx \tau_m \dot{\gamma}_c = 2$ which is unrealistically small. This suggests that the shear-thinning viscosity of such solutions can not be rationalised with the FENE-P model when choosing a relaxation time from filament thinning measurements. 

As shown in figure \ref{fig:h012_tau_varyc}(a), the apparent CaBER relaxation time $\tau_e$ increases in the dilute regime $c<c^*$ and, for the most dilute solutions, is less than the Zimm relaxation time calculated using
\begin{equation}
\tau_Z = \frac{1}{\zeta(3 \nu)} \frac{[\eta] M_w \eta_s}{N_a k_B T},
\label{eq:zimm}
\end{equation}
\noindent where $N_a$ is the Avogadro number, $k_B$ the Boltzmann constant, $T$ the temperature, $\zeta$ the Riemann zeta function and $\nu$ the solvent quality exponent. We used $\nu = 0.55$ between theta and good solvent to find $\tau_Z = 2.0$~ms. Similar results were reported by Clasen et al.  \cite{clasen2006dilute} who argue that the increase of $\tau_e$ in the dilute regime is caused by a self-concentration effect where chains start to interact while unravelling well beyond their equilibrium size in strong extensional flows, as was later rationalised by Prabhakar et al. \cite{prabhakar2016influence}. Clasen et al. and Prabhakar et al., who only considered cases where inertia was negligible in the Newtonian regime, also show that values of $\tau_e < \tau_Z$ arise at low polymer concentration where elasticity is too weak to fully overcome the solvent viscosity in the elastic regime. This effect should however be negligible for the aqueous PEO solutions of figure \ref{fig:h012_tau_varyc}(a) since it is inertia (and not the solvent viscosity) that dominates in the Newtonian regime (see figure \ref{fig:ht_varyc}(b)). We show in \S\ref{sec:FENE-P prediction for $h_1$} that values of $\tau_e < \tau_Z$ at low concentrations are consistent with polymer chains approaching their finite extensibility limit at the onset of the elastic regime (as anticipated by Campo-Deaño \& Clasen \cite{campo2010slow}), a case where equation \ref{eq:exponential} is no longer valid and filament thinning rates $\vert \dot{h}/h \vert > 1/3\tau$ are to be expected as we show in our previous paper \cite{gaillard2023beware}.

%Polymer chains should indeed be increasingly extended at the onset of the elastic regime as $c$ decreases since $h_1$ decreases (see figures \ref{fig:ht_varyc}(b) and \ref{fig:h012_tau_varyc}(b)), ultimately approaching full extension at sufficiently low concentrations as anticipated by Campo-Deaño \& Clasen \cite{campo2010slow}. 

%Note that using larger plate diameters would have lead to larger values of $\tau_e$, up to $\tau_m$, potentially above $\tau_Z$ which, if $\tau_m$ is the `real' relaxation time, would solve the apparent contradiction. 

%%%%%%%%%%%%%%%%%%%%%%%%%%%%%%%%%%%%%%%%%%%%%%%%%%%%%%%%%%%%%%%%%%%%%%%%%%%%%%%%%%%%%%%%%%%%%%%%%%%%%%%%%%%%%%%%%%%%%%%%%%%%%%%%
\section{Oldroyd-B prediction for $h_1$}
\label{sec:Oldroyd-B prediction for $h_1$}

In this section, we expand the polymer-relaxation-free theory leading to equation \ref{eq:h1_norelaxation} for $h_1$ to cases where polymer relaxation is not negligible in the Newtonian regime. The generalised Oldroyd-B prediction for $h_1$ is derived theoretically in \S\ref{subsec:Oldroyd-B Theory}, tested against experimental results in \S\ref{subsec:Oldroyd-B Experiments} and validated numerically in \S\ref{subsec:Oldroyd-B Simulations}.

\subsection{Theory}
\label{subsec:Oldroyd-B Theory}

The filament radius $h_1=h(t_1)$ marks the transition between the Newtonian regime ($t<t_1$), where the driving capillary force is balanced by inertia and/or viscosity, and the elastic regime ($t>t_1$) where capillarity is balanced by elastic stresses arising from the stretching of polymer chains. If inertia is negligible, slender filament theory predicts that the total (unknown) tensile force $T$ is constant along the liquid column (bridge or filament) \cite{eggers1997nonlinear,clasen2006beads} which, neglecting gravity and axial curvature effects, results in the zero-dimensional force balance equation \cite{clasen2006dilute}
\begin{equation}
(2X-1) \frac{\gamma}{h} = 3 \eta_s \dot{\epsilon} + \sigma_{p,zz} - \sigma_{p,rr}
\label{eq:force_balance}
\end{equation}
\noindent for a column of radius $h$. The driving capillary pressure $\gamma/h$ is balanced by the normal stress difference $\sigma_{zz} - \sigma_{rr}$ which is the sum of the solvent viscous stress $3 \eta_s \dot{\epsilon}$ and of the polymeric stress $\sigma_{p,zz} - \sigma_{p,rr}$ where $\dot{\epsilon} = -2 \dot{h}/h$ is the extension rate, the dot standing for $\mathrm{d}/\mathrm{d}t$, and where $z$ is the direction of the flow. The ratio $X = T / 2 \pi \gamma h$ may vary over time, approaching $X=0.7127$ for a Newtonian fluid \cite{mckinley2000extract}, hence recovering equation \ref{eq:selffimilar_viscous} close to breakup, and approaching $X=3/2$ in the elastic regime for an Oldroyd-B fluid \cite{eggers2020self} (and not $1$ as originally proposed in \cite{entov1997effect}). When inertia is not negligible, Tirtaatmadja et al. suggested adding a term of the form $\frac{1}{2} \rho \dot{h}^2$ to equation \ref{eq:force_balance} \cite{tirtaatmadja2006drop}, from which the inertio-capillary scaling of equation \ref{eq:selffimilar_inviscid} is recovered. In the elastic regime ($t>t_1$), assuming that inertia and/or solvent viscosity has become negligible, and assuming that the axial stress dominates over the radial stress, i.e. $\vert \sigma_{p,rr}  \vert \ll \vert \sigma_{p,zz} \vert$, the force balance equation reduces to 
\begin{equation}
(2X-1) \frac{\gamma}{h} = \sigma_{p,zz}.
\label{eq:force_balance_2}
\end{equation}

The elastic regime starts when the polymeric axial stress $\sigma_{p,zz}$, which increases over time in the Newtonian regime as polymer chains are progressively stretched by the extensional flow in the thinning bridge, becomes of the order of the capillary pressure, say, when it is equal to fraction $p$ of the capillary pressure (Campo-Deaño \& Clasen chose $p=1/2$ \cite{campo2010slow}). We hence get that $p \gamma/h_1 = \sigma_{p,zz}(t=t_1)$ where, for simplicity, the prefactor $2X-1$ of order unity has been included in the prefactor $p$. To estimate $h_1$, we hence need to choose a constitutive equation to express the polymeric stress. Since our main goal is to understand the effect of polymer relaxation during the Newtonian regime on $h_1$ which, when negligible, leads to equation \ref{eq:h1_norelaxation} for a single-mode Oldroyd-B fluid, we choose to use this model for simplicity. Indeed, although we know that the Oldroyd-B model is unable to capture the system-size dependence of the apparent relaxation time $\tau_e$ discussed in our previous paper \cite{gaillard2023beware} (see also figure \ref{fig:ht_water_and_PEG30} and \ref{fig:h12_taue}(a)), it is not yet clear whether it is able capture $h_1$ or not. Finite extensibility effects on $h_1$ will be discussed in \S\ref{sec:FENE-P prediction for $h_1$} using the FENE-P model.

For an Oldroyd-B fluid with elastic modulus $G$, we have $\sigma_{p,zz} = G (A_{zz} - 1)$ where $A_{zz}$ is the normal part of the conformation tensor ${\bf A}$ which follows (see equation \ref{eq:miguel_FENEP} and \cite{wagner2015analytic})
\begin{equation}
\dot{A}_{zz} - 2 \dot{\epsilon} A_{zz} = - \frac{A_{zz}-1}{\tau},
\label{eq:Azz_1}
\end{equation}
\noindent where $\tau$ is the relaxation time. Since we are interested in the location of highest polymer extensions along the bridge, we use the expression of the extension rate $\dot{\epsilon} = -2 \dot{h}/h$ at the thinnest point to obtain
\begin{equation}
\dot{A}_{zz} + \frac{4 \dot{h}}{h} A_{zz} = - \frac{A_{zz}-1}{\tau}.
\label{eq:Azz_2}
\end{equation}
\noindent Some (yet unknown) time after the onset of capillary thinning of the liquid bridge, polymer chains will have stretched well beyond their equilibrium size, i.e. $A_{zz} \gg 1$, at which point equation \ref{eq:Azz_2} can be integrated into 
\begin{equation}
A_{zz} h^4 \propto e^{-t/\tau},
\label{eq:Azz_3}
\end{equation}
\noindent with a (yet unknown) constant prefactor. Combining equations \ref{eq:Azz_3} and \ref{eq:force_balance_2} leads to the exponential decay $h \propto e^{-t/3\tau}$ described by equation \ref{eq:exponential}. Polymer chains are expected to remain close to their equilibrium coiled size ($A_{zz}$ close to $1$) until the extension rate in the thinning bridge approaches the coil-stretch transition value $1/2\tau$ predicted by the Oldroyd-B model. Beyond this point, following Clasen et al. \cite{clasen2009gobbling,campo2010slow}, we assume that polymer chains unravel with negligible relaxation, i.e., that the right-hand side of equation \ref{eq:Azz_2} becomes negligible so that $A_{zz} h^4$ becomes constant. More precisely,
\begin{equation}
A_{zz} h^4 = H^4,
\label{eq:Azz_4}
\end{equation}
\noindent where $H$ is the (yet unknown) bridge radius at which relaxation becomes negligible, which should correspond to the coil-stretch transition point at which $A_{zz}$ starts to become significantly larger than $1$. In particular, at the transition to the elastic regime at time $t=t_1$,
\begin{equation}
A_1 = (H/h_1)^4,
\label{eq:Azz_5}
\end{equation}
\noindent where $A_1 = A_{zz}(t_1)$ quantifies the amount of polymer stretching at the onset of the elastic regime. Since $p \gamma/h_1 = \sigma_{p,zz}(t_1) = G A_1$, we finally get that
\begin{equation}
h_1 = \left( \frac{G H^4}{p \gamma} \right)^{1/3},
\label{eq:h1_H}
\end{equation}
\noindent which is different from equation \ref{eq:h1_norelaxation}. Indeed, $H$ is only equal to $h_0$ in the limit where polymer relaxation is negligible throughout the whole Newtonian regime so that $A_{zz} h^4$ is constant and equal to $h_0^4$ since $A_{zz} = 1$ at the onset of capillary thinning, assuming no pre-stress. In other words, $H=h_0$ means that the coil-stretch transition starts at the onset of capillary thinning. This is only true if the relaxation time $\tau$ is much larger than the time taken by the liquid bridge to thin from $h_0$ to $h_1$, as we detail below.

\subsection{Experiments}
\label{subsec:Oldroyd-B Experiments}

In order to test the validity of the Oldroyd-B prediction (equation \ref{eq:h1_H}) for $h_1$, we first need to compute $H$ from the time-evolution of $A_{zz}$. Note that we do not experimentally measure the extension of polymer chains, unlike Ingremeau \& Kellay \cite{ingremeau2013stretching} who confirmed the transition from a coiled to a stretched state in viscoelastic pinch-off using fluorescently labelled DNA. Rather, since our goal is to test a specific constitutive equation, here Oldroyd-B, we calculate its prediction for $A_{zz}(t)$ using equation \ref{eq:Azz_2} where the extension rate $\dot{\epsilon} = -2 \dot{h}/h$ is taken from experimental values of $h(t)$. In other words, we calculate the prediction of the model for the experimental history of extension rates in the bridge / filament. In particular, we do \emph{not} assume large polymer extension ($A_{zz} \not\gg 1$) since the point at which $A_{zz}$ starts to become significantly larger than $1$ is precisely what sets $H$. Equation \ref{eq:Azz_2} can in fact be integrated, as shown by Bazilevsky et al. \cite{bazilevsky2001breakup}, introducing a function $y(t)$ such that $A_{zz} = y \exp{(-t/\tau)} / h^4$, which leads to $\dot{y} = h^4 \exp{(t/\tau)} / \tau$, yielding 
\begin{equation}
A_{zz} = \frac{e^{-t/\tau}}{h^4} \left( h_0^4 \, e^{t_0/\tau} + \frac{1}{\tau} \int_{t_0}^t h^4(t')\,  e^{t'/\tau} \mathrm{d} t' \right),
%A_{zz} = \frac{\exp{(-t/\tau)}}{h^4} \left( h_0^4 \exp{(t_0/\tau)} + \frac{1}{\tau} \int_{t_0}^t h^4(t') \exp{(t'/\tau)} \mathrm{d} t' \right)
\label{eq:solution_Bazilevsky}
\end{equation}
\noindent where the initial time $t_0$ corresponds to the onset of capillary thinning, i.e. $h(t_0) = h_0$ and $A_{zz}(t_0)=1$ (no pre-stress). Since the $h(t)$ history is set by the experimental data, the only adjustable parameter of equation \ref{eq:solution_Bazilevsky} is the relaxation time $\tau$. In the following, we either use the apparent ($\tau_e$) or the maximum ($\tau_m$) relaxation time measured experimentally (see figure \ref{fig:h12_taue}(a)) to calculate $A_{zz}$ since we still do not know which one is the `true' one, if any.

\begin{figure}[t!]
  \centerline{\includegraphics[scale=0.58]{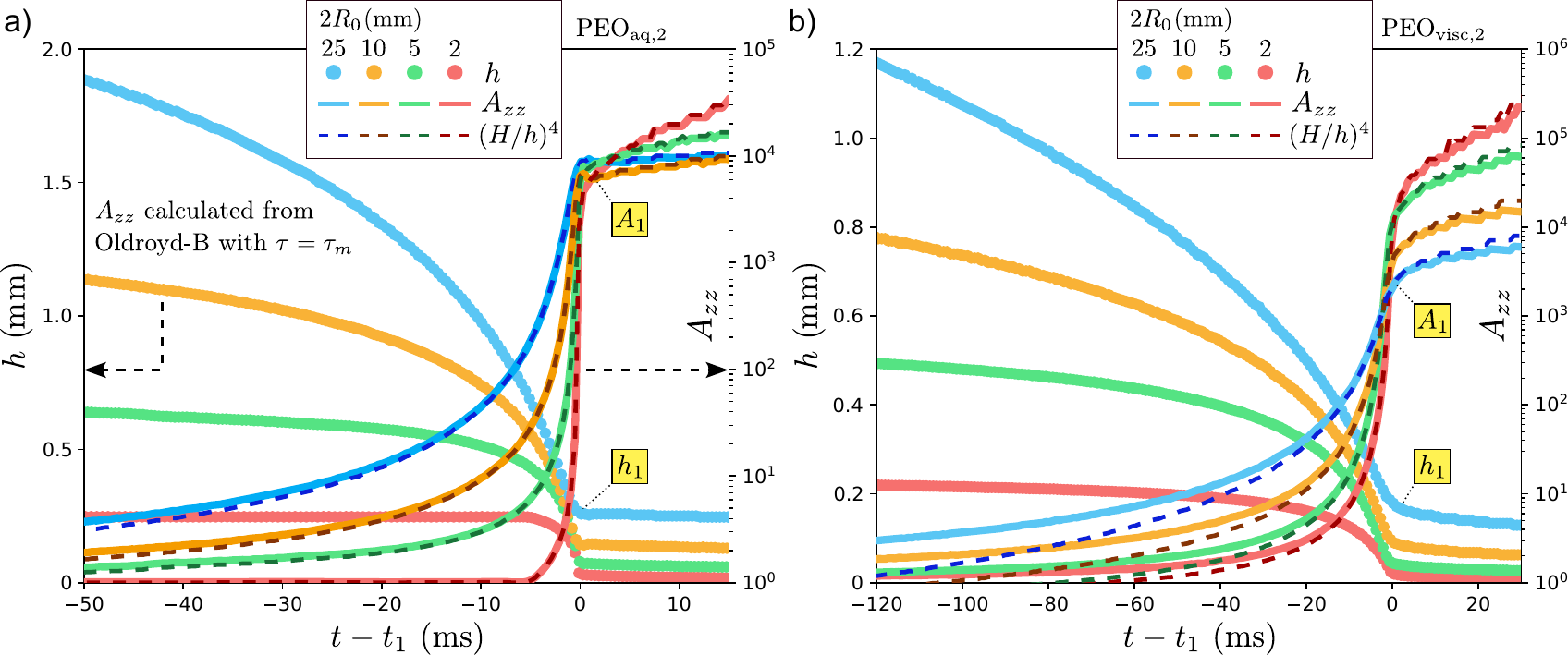}}
  \caption{Time evolution of the experimental minimum filament / bridge radius $h$ and of $A_{zz}$, calculated from the Oldroyd-B prediction (equation \ref{eq:solution_Bazilevsky}) using the experimental values of $h(t)$ with the choice of relaxation time $\tau = \tau_{m}$ (see figure \ref{fig:h12_taue}(a)), for plate diameters $2R_0 = 2$, $5$, $10$ and $25$~mm for the PEO$_{\mathrm{aq,2}}$ (a) and the PEO$_{\mathrm{visc,2}}$ (b) solutions. Time $t_1$ marks the onset of the elastic regime, with $h_1 = h(t_1)$ and $A_1 = A_{zz}(t_1)$. Values of $A_{zz}$ in the Newtonian regime ($t<t_1$) are compared to $(H/h)^4$ (see equation \ref{eq:Azz_4}) where $H$ is used as a fitting parameter to optimise the agreement close to $t_1$.}
\label{fig:Azz_Baz}
\end{figure}

Values of $A_{zz}(t)$ computed from equation \ref{eq:solution_Bazilevsky} using the experimental values of $h(t)$ with relaxation time $\tau = \tau_m$ are shown in figure \ref{fig:Azz_Baz}(a) for the PEO$_{\mathrm{aq,2}}$ solution and in figure \ref{fig:Azz_Baz}(b) for the PEO$_{\mathrm{visc,2}}$ solution for plate diameters $2R_0$ between $2$ and $20$~mm. The experimental values of $h(t)$ are shown on the left y-axis and the time reference $t_1$ corresponds to the onset of the elastic regime. We find that the amount of polymer extension $A_1 = A_{zz}(t_1)$ at the onset of the elastic regime is fairly independent of the initial condition for the PEO$_{\mathrm{aq,2}}$ while, for the PEO$_{\mathrm{visc,2}}$ solution, $A_1$ decreases as $h_0(R_0,V^*)$ increases, as mentioned in our previous paper \cite{gaillard2023beware} and as we are about to discuss here in more depth. In any case, we find that $A_{zz}$ always increases as $1/h^4$ in the Newtonian regime close enough to the transition to the elastic regime. More specifically, values of $A_{zz}$ are well captured by $(H/h)^4$ using $H$ as a fitting parameter for each data set (close to $t_1$), from which $H$ is estimated, see figure \ref{fig:Azz_Baz}.

\begin{figure}
  \centerline{\includegraphics[scale=0.58]{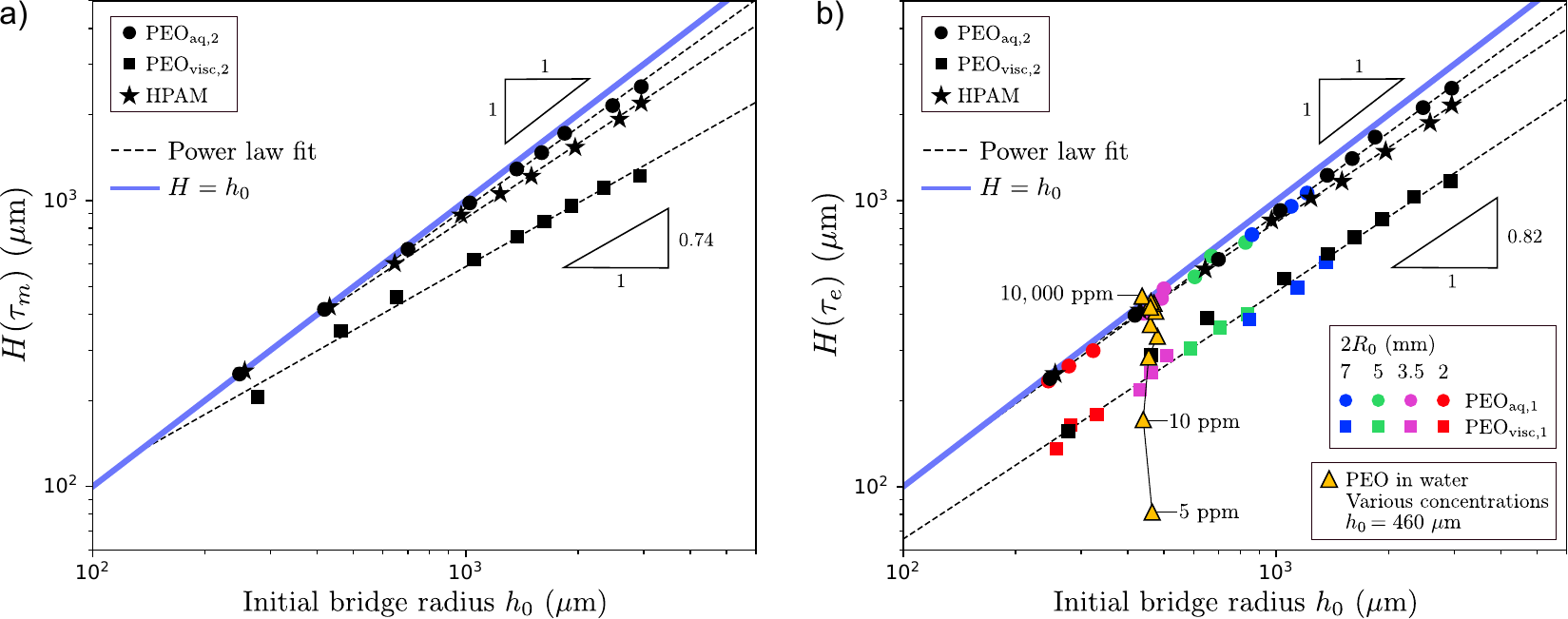}}
  \caption{Values of $H$ calculated from the time evolution of $A_{zz}$ from Oldroyd-B (see figure \ref{fig:Azz_Baz}) using the maximum relaxation time $\tau=\tau_m$ (a) or the effective relaxation time $\tau=\tau_e$ (b) for different polymer solutions and initial bridge radii $h_0(R_0,V^*)$, plotted against $h_0$. The line $H=h_0$ is shown in both panels.}
\label{fig:H}
\end{figure}

Values of $H$ calculated using $\tau = \tau_m$, named $H(\tau_m)$, are plotted against $h_0$ in figure \ref{fig:H}(a) for the PEO$_{\mathrm{aq,2}}$, PEO$_{\mathrm{visc,2}}$ and HPAM solutions which are the only solutions for which sufficiently large plate diameters were used to estimate the maximum relaxation time $\tau_m$ (the high-$h_0$ limit of $\tau_e$, see figure \ref{fig:h12_taue}(a)). For the PEO$_{\mathrm{aq,2}}$ and HPAM solutions, we find that $H$ is essentially equal to $h_0$ at low $h_0$ and that $H/h_0$ decreases as $h_0$ increases, down to $0.73$ for the largest plate diameter. In contrast, for the PEO$_{\mathrm{visc,2}}$ solution, the ratio $H/h_0$ takes significantly smaller values, decreasing from $0.56$ to $0.40$ as $h_0$ increases. This is why the $(H/h)^4$ fit for $A_{zz}$ is fairly good throughout the whole Newtonian regime for the PEO$_{\mathrm{aq,2}}$ solution in figure \ref{fig:Azz_Baz}(a) while it is only valid within a small time window close to the transition to the elastic regime for the PEO$_{\mathrm{visc,2}}$ solution in figure \ref{fig:Azz_Baz}(b). Indeed, if $H=h_0$, then the $(H/h)^4$ fit for $A_{zz}$ is even valid at the onset of capillary thinning where $h=h_0$ and $A_{zz} = 1$.

Figure \ref{fig:H}(a) hence suggests that, although all three solutions have comparable relaxation times (see figure \ref{fig:h12_taue}(a)), the thinning dynamics in the Newtonian regime is such that polymer chains stretch almost without relaxing in the Newtonian regime for the low solvent viscosity solutions (PEO$_{\mathrm{aq,2}}$ and HPAM) while relaxation is not negligible for the high solvent viscosity (PEO$_{\mathrm{visc,2}}$) solution. This is because the Newtonian thinning dynamics is slower for the most viscous solution, see, e.g., figure \ref{fig:Azz_Baz} where, for $2R_0 = 2$~mm, the bridge takes only about $6$~ms to thin from $h_0$ to $h_1$ for the PEO$_{\mathrm{aq,2}}$ solution, much less than $\tau_m$ (chains do not have enough time to relax), while it takes about $900$~ms for the PEO$_{\mathrm{visc,2}}$, much more than $\tau_m$ (not visible in figure \ref{fig:Azz_Baz}(b) where we focus on times close to $t_1$). The fact that $H/h_0$ increases as $h_0$ increases can therefore be interpreted by a longer time to thin from $h_0$ to $h_1$ as $h_0$ increases, consistent with the fact that the Rayleigh and viscous time scales $\tau_R = \sqrt{\rho h_0^3 / \gamma}$ and $\tau_{\mathrm{visc}} = \eta_0 h_0 / \gamma$ both increase with $h_0$.

The dependence of $H$ on $h_0$ and on the solvent viscosity explains why, in figure \ref{fig:Azz_Baz} and in our previous paper \cite{gaillard2023beware}, $A_1$ is independent of $h_0$ for low solvent viscosity solutions (PEO$_{\mathrm{aq,2}}$ and HPAM) while $A_1$ decreases as $h_0$ increases for the high solvent viscosity solution (PEO$_{\mathrm{visc,2}}$). For the former where $H$ is close to $h_0$ (negligible polymer relaxation in the Newtonian regime), $A_1 = (H/h_1)^4$ (see equation \ref{eq:Azz_5}) is close to $(h_0/h_1)^4$ which is fairly constant since $h_1 \propto h_0$ according to figure \ref{fig:h12_taue}(b). For the latter however, where $H \propto h_0^{0.74}$ in our range of $h_0$ according to figure \ref{fig:H}(a), we get $A_1 = (H/h_1)^4 \propto h_0^{-1.04}$.

In figure \ref{fig:H}(b) we plot the values of $H$, named $H(\tau_e)$, calculated using the apparent relaxation time $\tau = \tau_e$ when computing $A_{zz}$ from equation \ref{eq:solution_Bazilevsky} (instead of its large-$h_0$ limit $\tau_m$ in figure \ref{fig:H}(a)). All polymer solutions are now featured since $\tau_e$ is measured for any experiment from an exponential fit of $h(t)$ in the exponential part of the elastic regime (see figure \ref{fig:ht_water_and_PEG30}), i.e., the PEO$_{\mathrm{aq,2}}$, PEO$_{\mathrm{visc,2}}$ and HPAM solutions, like in figure \ref{fig:H}(a), but also the PEO$_{\mathrm{aq,1}}$ and PEO$_{\mathrm{visc,1}}$ solutions, where three different non-dimensional droplet volumes $V^*$ were tested for each of the four smallest plate diameters, as well as the PEO solutions in water with various polymer concentration (table \ref{tab:rheology_PEO}) which were tested for a single $(R_0,V^*)$ set corresponding to $h_0 \approx 460$~$\mu$m. The data corresponding to the PEO solutions with different polymer concentrations $c$ show how, for a given flow history in the Newtonian regime (same $h(t)$ curves for $t<t_1$ for all concentrations, see figure \ref{fig:ht_varyc}), $H$ increases with $c$ via the increase in the relaxation time (here $\tau = \tau_e$), reaching the upper limit value $h_0$ at large $\tau$. This is because, for large $\tau$ values, polymer relaxation is negligible throughout the whole Newtonian regime while, for low $\tau$ values, $A_{zz}$ remains equal to $1$ for most of the Newtonian regime, only increasing when $\dot{\epsilon}$ finally gets of the order of $1/2\tau$ close to the transition to the elastic regime. %Consequently, values of $H$ are smaller in figure \ref{fig:H}(b) than in figure \ref{fig:H}(a) for the PEO$_{\mathrm{aq,2}}$, PEO$_{\mathrm{visc,2}}$ and HPAM solutions since $\tau_e \le \tau_m$, with a greater ratio $\tau_m/\tau_e$ at low $h_0$ values (see figure \ref{fig:h12_taue}(a)) and hence a greater $H(\tau_m)/H(\tau_e)$ ratio.

%The previous discussions on $H$ suggest that the ratio $H/h_0$ is a function of a Deborah number $\Deb = \tau / \tau_N$ where $\tau$ is the polymer relaxation time and $\tau_N$ is the `Newtonian time scale', i.e. the time scale of the thinning dynamics in the Newtonian regime, which should scale as the Rayleigh time scale $\tau_R = \sqrt{\rho h_0^3 / \gamma}$ or as the viscous time scale $\tau_{\mathrm{visc}} = \eta_0 h_0 / \gamma$ depending on the Ohnesorge number $\Oh$, see equation \ref{eq:Oh}, with $H/h_0 = 1$ for $\Deb \gg 1$. This will be confirmed and discussed further in \S\ref{sec:Numerical results on $h_1$} using numerical simulations.

Now that we know the value of $H$, we can test the validity of equation \ref{eq:h1_H} for the filament radius $h_1$ at the onset of the elastic regime. The value of the elastic modulus, $G = \eta_p/\tau$ in the Oldroyd-B model, is however not uniquely defined since, while $\eta_p = \eta_0 - \eta_s$ can be calculated unambiguously from the shear rheology, the relaxation time $\tau$ could either be the apparent one $\tau_e$ or the maximum one $\tau_m$ since we do not know yet which one is the `true' one, if any. We hence need to test for both. To this end, we define 
\begin{equation}
G_H = \gamma h_1^3 / H^4,
\label{eq:GH_def}
\end{equation}
\noindent where $h_1$ is the value measured experimentally, which should be $G_H = G/p$ according to equation \ref{eq:h1_H}.

\begin{figure}
  \centerline{\includegraphics[scale=0.58]{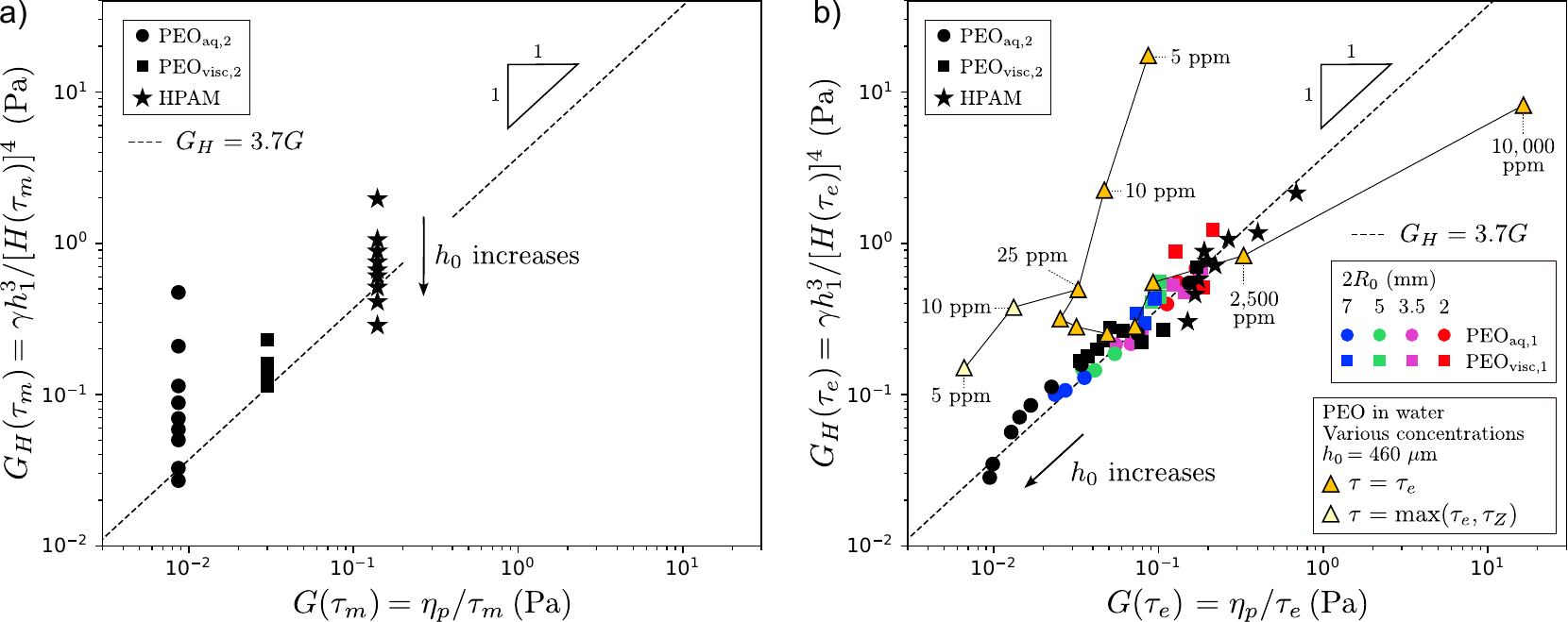}}
  \caption{$G_H$ defined in equation \ref{eq:GH_def} against the elastic modulus $G = \eta_p/\tau$ where $H$ and $G$ are calculated from the maximum relaxation time $\tau = \tau_m$ (a) or the effective relaxation time $\tau = \tau_e$ (b) for various polymer solutions and initial bridge radii $h_0(R_0,V^*)$. Values of $h_1$ are the one measured experimentally. In panel (b), for the aqueous PEO solutions of different concentrations (see table \ref{tab:rheology_PEO}), we show the effect replacing $\tau_e$ by the Zimm relaxation time $\tau_Z$ when calculating $G$ and $H$ for the lowest concentrations $c = 5$ and $10$~ppm which exhibit values of $\tau_e < \tau_Z$, see figure \ref{fig:h012_tau_varyc}(a). The line $G_H = 3.7 G$ is shown in both panels.}
\label{fig:GH}
\end{figure}

In figure \ref{fig:GH}(a), $G_H$ is plotted against $G$ for the choice of relaxation time $\tau = \tau_m$ for the PEO$_{\mathrm{aq,2}}$, PEO$_{\mathrm{visc,2}}$ and HPAM solutions which are the only solutions for which sufficiently large plate diameters were used to estimate $\tau_m$. More precisely, values of $G_H(\tau_m)$ calculated from $H(\tau_m)$ are plotted against $G(\tau_m) = \eta_p / \tau_m$ which takes a unique value for each solution since the relaxation time is unique. We find that values of $G_H(\tau_m)$ are, however, not unique for the PEO$_{\mathrm{aq,2}}$ and HPAM solutions and monotonously decrease as $h_0$ is increases. For the PEO$_{\mathrm{visc,2}}$ however, values of $G_H$ vary between $0.11$ and $0.23$ without clear monotonic trend as $h_0$ increases. This is because, since $h_1 \propto h_0$ and $H \propto h_0^{k_H}$ in the range of $h_0$ values investigated with $k_H \le 1$ (see figures \ref{fig:h12_taue}(b) and \ref{fig:H}(a)), $G_H \propto h_1^3 / H^4 \propto h_0^{3-4k_H}$ which means that $G_H$ is only expected to be independent of $h_0$ for $k_H = 0.75$, very close to the value $0.74$ found for the PEO$_{\mathrm{visc,2}}$ solution in figure \ref{fig:H}(a). This suggests that the prediction of equation \ref{eq:h1_H} for the choice $\tau = \tau_m$ is only potentially valid for one of our three solutions, the most dilute and viscous one.

By contrast, as shown in figure \ref{fig:GH}(b), when choosing $\tau_e$ instead of $\tau_m$ for the relaxation time to calculate $G_H(\tau_e)$ from $H(\tau_e)$ and $G(\tau_e) = \eta_p/\tau_e$, data points fall on a single curve for all three solutions (PEO$_{\mathrm{aq,2}}$, PEO$_{\mathrm{visc,2}}$ and HPAM) as well as for the PEO$_{\mathrm{aq,1}}$ and PEO$_{\mathrm{visc,1}}$ solutions (for which $V^*$ is varied for the four smallest plate diameters) and we find the linear relationship $G_H = G/p$ predicted by equation \ref{eq:h1_H} with $p \approx 0.27$. It is quite remarkable that the Oldroyd-B model, derived for ideal dilute chains, is able to capture the transition to the elastic regime for solutions that are as diverse in terms of solvent viscosity, concentration and, potentially, solvent quality exponents.

%The only thing that these PEO$_{\mathrm{aq}}$ (1 and 2), PEO$_{\mathrm{visc}}$ (1 and 2) and HPAM solutions have in common is comparable relaxation times (see figure \ref{fig:h12_taue}(a)) and equation \ref{eq:h1_H} should, in principle, be valid for any relaxation time. However,

Strong deviations from the $G_H = 3.7 G$ line can be observed in figure \ref{fig:GH}(b) at low polymer concentrations for the data corresponding to the PEO solutions with various polymer concentrations. This data set can be broken down into three subsets: for typically $c < 100$~ppm, $G_H$ decreases sharply with concentration while is it almost constant for $100$~ppm$ \le c < 2,500$~ppm and increases sharply with concentration for $c \ge 2,500$~ppm. These trends can be explained by the fact that $G_H \propto h_1^3/H^4$ where $h_1$ increases very slowly with concentration as $h_1 \propto c^{0.16}$ for $c<2,500$~ppm and increases sharply for higher concentrations (see figure \ref{fig:h012_tau_varyc}(b)) while $H$ increases sharply with concentration for typically $c < 100$~ppm, becoming almost constant and equal to $h_0$ for larger concentrations (see figure \ref{fig:H}(b)). These strong deviations from the $G_H = 3.7 G$ line observed at low polymer concentrations in figure \ref{fig:GH}(b) could be partially explained by the fact that apparent relaxation times (measured from exponential fitting of $h(t)$) are less than the Zimm relaxation time $\tau_Z = 2$~ms for $c = 5$ and $10$~ppm (see figure \ref{fig:h012_tau_varyc}(a)). In figure \ref{fig:GH}(b), we show the effect of choosing $\tau = \tau_Z$ instead of $\tau_e$ as the relaxation time for $c = 5$ and $10$~ppm, which changes the value of both $G_H$ via $H(\tau)$ and of $G = \eta_p/\tau$. We find that this correction leads to data points significantly closer to the  $G_H = 3.7 G$ line, mainly stemming from larger values of $H$ than in figure \ref{fig:H}(b), although one order of magnitude deviation from the $G_H = 3.7 G$ line still remains. We show in \S\ref{sec:FENE-P prediction for $h_1$} that finite extensibility effects can explain this deviation (i.e. values of $h_1$ higher than the Oldroyd-B prediction) as well as the values of $\tau_e < \tau_Z$ for low polymer concentrations.

The large deviation from the $G_H = 3.7 G$ line for the entangled $10,000$~ppm solution in figure \ref{fig:GH}(b) (the only solution for which polymers affect the thinning dynamics even before the exponential regime, see figure \ref{fig:ht_varyc}), is probably due to the fact that such solutions cannot be described by non-interacting polymer theories such as Oldroyd-B.

In conclusion, when polymer relaxation is not negligible in the Newtonian regime ($t<t_1$), equation \ref{eq:h1_norelaxation} should be replaced by equation \ref{eq:h1_H} which gives $h_1 \propto H^{4/3}$ where $H \le h_0$. In our experiments, the power law dependence of $H$ on $h_0$ (see figure \ref{fig:H}) leads to the fairly proportional relationship between $h_1$ and $h_0$ observed in figure \ref{fig:h12_taue}(b), differences in slopes among different liquids stemming from differences in elastic moduli $G$. In \S\ref{subsec:Oldroyd-B Simulations}, we discuss how $H$ scales with the parameters of the problem using numerical simulations.

\subsection{Simulations}
\label{subsec:Oldroyd-B Simulations}

%Note that this section is not about the system-size dependence of the apparent relaxation time $\tau_e$ since, in the Oldroyd-B limit ($L^2 \to \infty$), the filament thinning rate in the elastic regime is constant and equal to $1/3\tau$ independently of the plate diameter or droplet volume. The possibility of explaining variations of $\tau_e$ by finite-extensibility effects have been explored in our previous paper \cite{gaillard2023beware}. 

To further investigate the effect of polymer relaxation on the transition radius $h_1$ marking the onset of the elastic regime, we now consider numerical simulations using the Oldroyd-B model ($L^2 = + \infty$) with a single relaxation time $\tau$ (finite extensibility effects will be discussed in \S\ref{sec:FENE-P prediction for $h_1$}). In this paper, numerical simulations are not directly compared to experiments but are rather used to validate theoretical expressions which are then compared with experiments (see our previous paper for direct experiment-simulation comparisons \cite{gaillard2023beware}).

In order capture $h_1$, we first need to capture the critical bridge radius $H$ at which polymer relaxation (the right-hand side of equations \ref{eq:Azz_1} or \ref{eq:miguel_FENEP}) becomes negligible in the Newtonian regime ($t<t_1$). $H$ is the bridge radius marking the onset of the coil-stretch transition at which polymer chains start to extend significantly beyond their equilibrium shape, i.e., at which $A_{zz}$ start becoming significantly larger than $1$, see \S\ref{subsec:Oldroyd-B Theory}. We already know that $H \to h_0$ in the limit where the relaxation time $\tau$ is so large that polymer relaxation is always negligible in the Newtonian regime, a limit where equation \ref{eq:h1_H} reduces to the classical formula of equation \ref{eq:h1_norelaxation}. The goal of this subsection is therefore to expand our knowledge to cases where relaxation is not negligible in the Newtonian regime using the Oldroyd-B model.

%$H$ is indeed the key to predict the bridge radius $h_1$ marking the onset of the elastic regime where elasticity (building up during the Newtonian regime without yet influencing its dynamics) finally overcomes inertia and/or viscosity. 

\begin{figure}
  \centerline{\includegraphics[scale=0.58]{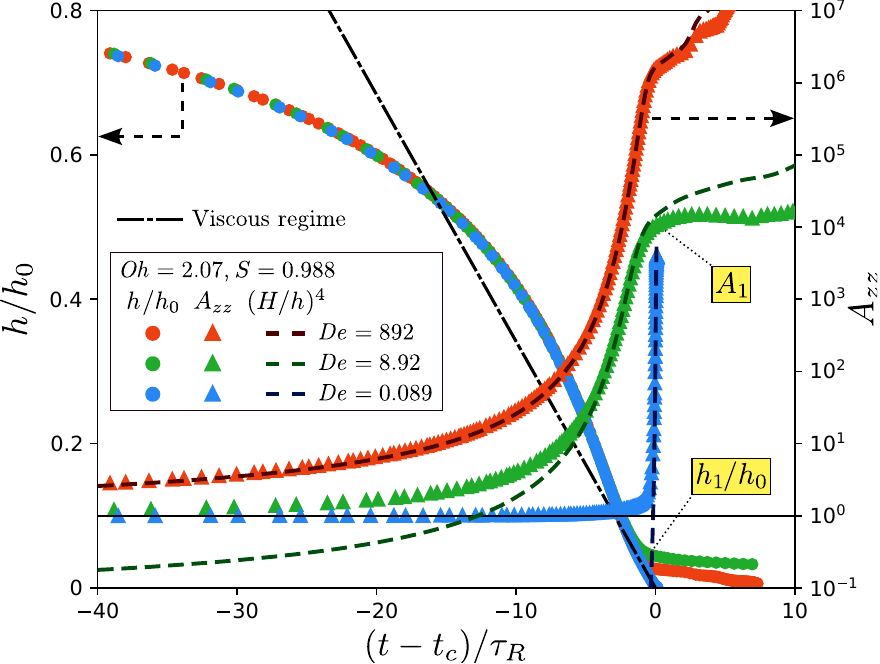}}
  \caption{Numerical time evolution of the non-dimensional (minimum) bridge / filament radius $h/h_0$ and of the (maximum) polymer extension $A_{zz}$ for $\Oh = 2.07$, $S=0.988$ and $h_0/R_0 = 0.23$ and three different Deborah numbers. $A_{zz}$ is compared with $(H/h)^4$ where $H$ is used as a fitting parameter. The self-similar viscous regime of equation \ref{eq:selffimilar_viscous}, or equivalently $h/h_0 = 0.0709 (t_c-t)/(\Oh \tau_R)$, is also plotted, where $t_c$ is the time at which the filament would break if the transition to an elastic regime, at $h=h_1$, did not occur.}
\label{fig:num_h_Azz}
\end{figure}

Figure \ref{fig:num_h_Azz} shows the numerical time evolution of the non-dimensional minimum bridge / filament radius $h/h_0$ for a fixed Ohnesorge number $\Oh = 2.07$, viscosity ratio $S = 0.988$ and $h_0/R_0 = 0.23$ with three Deborah numbers $\Deb$ spanning four orders of magnitude (see equations \ref{eq:Oh}, \ref{eq:De} and \ref{eq:S} for definitions). We only consider the value of $A_{zz}$ at this minimum-radius position along the bridge / filament since this is where polymer chains are the most stretched. This maximum value, simply denoted $A_{zz}$ from now on, is plotted in figure \ref{fig:num_h_Azz} on the right $y$-axis. Note that all $h/h_0$ curves are identical in the Newtonian regime, only diverging at the transition to the elastic regime at different radii $h_1$. The time reference $t_c$ corresponds to the time at which the bridge would break if this transition did not occur. This is highlighted by the fact that the self-similar viscous thinning law of equation \ref{eq:selffimilar_viscous}, which becomes $h/h_0 = 0.0709(t_c-t)/(\Oh \tau_R)$ with our choice of non-dimensionalisation and which is plotted in figure \ref{fig:num_h_Azz}, fits numerical results close to the transition. Note that simulations could often not be continued long after the transition.

Figure \ref{fig:num_h_Azz} shows how, for a given flow history in the Newtonian regime, polymer chains start stretching at different times depending on the Deborah number. For $\Deb \gg 1$, relaxation is always negligible in the Newtonian regime and $A_{zz}$ therefore increases as $(H/h)^4$ where $H=h_0$, see discussion in \S\ref{subsec:Oldroyd-B Theory}. For $\Deb \ll 1$ however, the flow only becomes strong enough to start stretching polymers (beyond their equilibrium shape) at small bridge radii where the thinning dynamics has become self-similar and, for $\Oh \gg 1$, follows equation \ref{eq:selffimilar_viscous}. In that case, $A_{zz} = 1$ for most of the Newtonian regime, only increasing close to the transition to the elastic regime, following $A_{zz} = (H/h)^4$ where $H \ll h_0$ is the characteristic bridge radius at which $A_{zz}$ starts increasing. Values of $H$ are estimated by fitting $A_{zz}$ with $(H/h)^4$, using $H$ as a fitting parameter, see figure \ref{fig:num_h_Azz}, as was done in figure \ref{fig:Azz_Baz} for experimental results.

\begin{figure}[t!]
  \centerline{\includegraphics[scale=0.58]{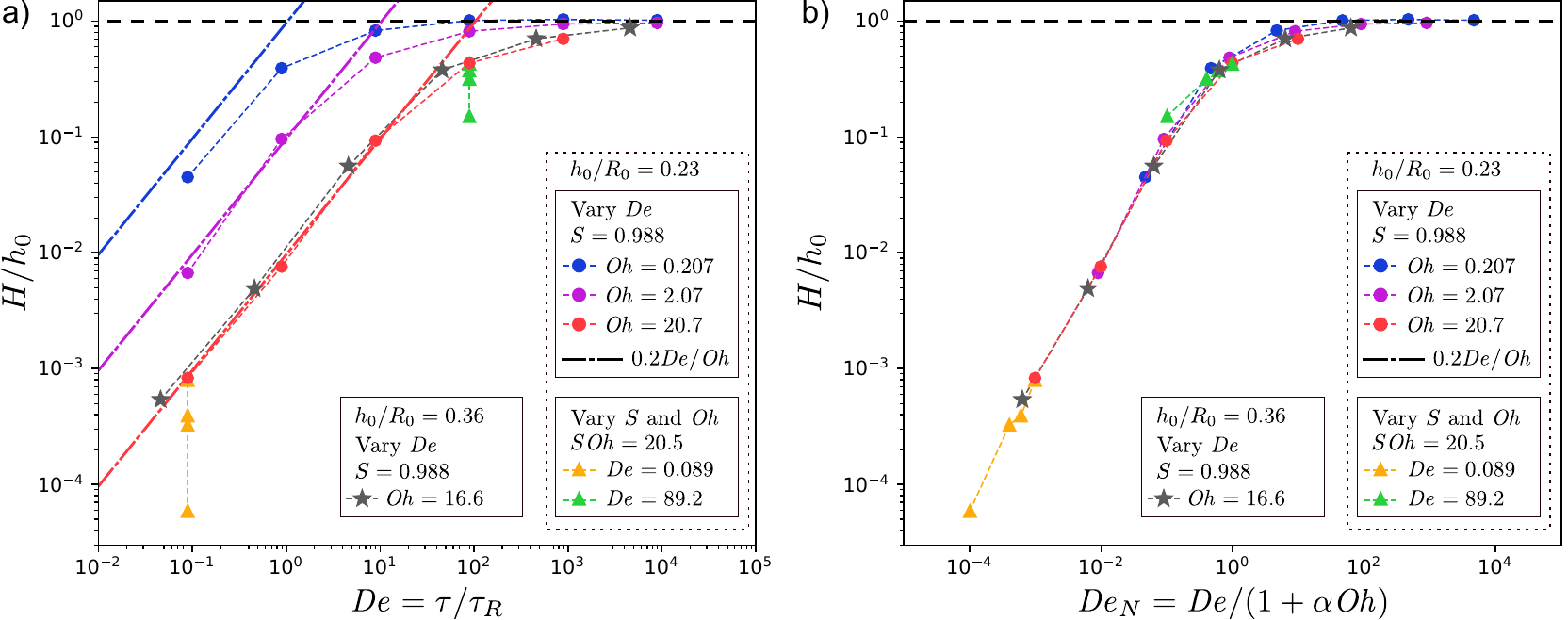}}
  \caption{Numerical values of $H/h_0$ against the Deborah number $\Deb = \tau/\tau_R$ based on the inertio-capillary time scale $\tau_R$ (a) and against the general Deborah number $\Deb_N = \tau/\tau_N$ based on the general time scale $\tau_N =  \tau_R (1 + \alpha \Oh)$ with $\alpha = 4.3$ (b) for various parameters, the same in (a) and (b). Dots ($\bullet$) correspond to $h_0/R_0 = 0.23$ and $S = 0.988$ with $\Oh = 0.207$ (blue) $2.07$ (purple) and $20.7$ (red) with $\Deb$ ranging between $8.92 \times 10^{-2}$ and $8.92 \times 10^{3}$ (last $\Deb$ excluded for the largest $\Oh$). Triangles ($\blacktriangle$) correspond to $h_0/R_0 = 0.23$ with $\Deb = 8.92 \times 10^{-2}$ (yellow) and $8.92 \times 10^{1}$ (green) with varying both $S$ (between $0.1$ and $0.988$) and $\Oh$ while keeping $S \Oh $ constant and equal to $9.88$. Stars ($\star$) correspond to $h_0/R_0 = 0.362$ with $\Oh = 16.6$ and $S = 0.988$ with $\Deb$ ranging between $4.59 \times 10^{-2}$ and $8.92 \times 10^{3}$.}
\label{fig:num_H_De}
\end{figure}

Values of $H/h_0$ are plotted against the Deborah number in figure \ref{fig:num_H_De}(a) for Ohnesorge numbers between $0.2$ and $20$. The low-$\Deb$ behaviour corresponds to cases where polymers only start stretching within the self-similar thinning regime where the thinning dynamics follows a scaling of the form $h = B (t_c-t)^{\beta}$ with $B \sim (\gamma / \rho)^{1/3}$ and $\beta = 2/3$ in the inviscid limit ($\Oh \ll 1$, see equation \ref{eq:selffimilar_inviscid}) and $B \sim \gamma / \eta_0$ and $\beta = 1$ in the viscous limit ($\Oh \gg 1$, see equation \ref{eq:selffimilar_viscous}). The coil-stretch transition occurs when the extension rate $\dot{\epsilon} = -2 \dot{h}/h \sim 2 \beta /(t_c-t)$ becomes of the order of $1/\tau$, i.e., at a time $t_H = t_c - 2 \beta \tau$. Therefore, the bridge radius $H = h(t_H)$ marking the onset of the coil-stretch transition scales as
\begin{equation}
H \sim (\gamma \tau^2 / \rho)^{1/3} \quad \Leftrightarrow \quad H/h_0 \sim \Deb^{2/3}
\label{eq:H_inviscid}
\end{equation}
\noindent in the inviscid limit ($\Oh \ll 1$), as first derived by Campo-Deaño \& Clasen \cite{campo2010slow}, or as 
\begin{equation}
H \sim \gamma \tau / \eta_0 \quad \Leftrightarrow \quad H/h_0 \sim \Deb/\Oh = \tau/\tau_{\mathrm{visc}}
\label{eq:H_viscous}
\end{equation}
\noindent in the viscous limit ($\Oh \gg 1)$. The scaling of equation \ref{eq:H_viscous} is shown in figure \ref{fig:num_H_De}(a) with a prefactor $0.2$ and shows a good agreement with the values of $H$ for the two largest Ohnesorge numbers. Unfortunately, no simulations could be performed for $\Oh \ll 1$ to test equation \ref{eq:H_inviscid}. Note that, in this peculiar limit where polymer chains only start stretching within the self-similar thinning regime, $H$, and therefore $h_1$ given by equation \ref{eq:h1_H}, do not depend on $h_0$ and are therefore independent of the size of the system, in sharp contrast with the high-$\Deb$ limit where $H=h_0$ and therefore $h_1 \propto h_0^{4/3}$, see equation \ref{eq:h1_norelaxation}. In fact, inserting equation \ref{eq:H_inviscid} or \ref{eq:H_viscous} into equation \ref{eq:h1_H} gives 
\begin{equation}
h_1 \sim \left(G (\gamma \tau^8 / \rho^4)^{1/3} \right)^{1/3}
\label{eq:h1_inviscid}
\end{equation}
\noindent in the inviscid limit ($\Oh \ll 1$), or
\begin{equation}
h_1 \sim (G \gamma^3 \tau^4 / \eta_0^4)^{1/3}
\label{eq:h1_viscous}
\end{equation}
\noindent in the viscous limit ($\Oh \gg 1)$.

In the high-$\Deb$ limit, $H/h_0 \to 1$ since polymer relaxation becomes negligible even at the onset of capillary thinning where $h=h_0$. However, while all curves in figure \ref{fig:num_H_De}(a) have the same shape, the Deborah number at which $H/h_0$ reaches $1$ depends on the Ohnesorge number. This is because we chose to express the Deborah number as $\Deb = \tau/\tau_R$ where $\tau_R = (\rho h_0^3 / \gamma)^{1/2}$ is the inertio-capillary time scale which is not relevant for the moderate to large Ohnesorge number featured in figure \ref{fig:num_H_De}(a). The relevant time scale for the thinning dynamics at large $\Oh$ is $\tau_{\mathrm{visc}} = \Oh \tau_R = \eta_0 h_0 / \gamma$ and we would hence expect that $H/h_0 = \mathcal{O}(1)$ not when $\tau/\tau_R = \mathcal{O}(1)$ but when $\tau/\tau_{\mathrm{visc}} \mathrm{(} = \Deb / \Oh \mathrm{)} = \mathcal{O}(1)$. In figure \ref{fig:num_H_De}(b), we show that values of $H/h_0$ indeed rescale on a single curve when plotted against a generalised Deborah number $\Deb_N = \tau/\tau_N$ where $\tau_N$, defined as 
\begin{equation}
\tau_N = \tau_R (1 + \alpha \Oh),
\label{eq:tauN}
\end{equation}
\noindent is an empirical attempt at expressing the general time scale of the thinning dynamics in the Newtonian regime for any $\Oh$, connecting the low and high-$\Oh$ scalings $\tau_R$ and $\tau_{\mathrm{visc}}$, where $\alpha = 4.3$ is a fitting parameter. This scaling ensures that $H/h_0 = \mathcal{O}(1)$ when $\Deb_N = \mathcal{O}(1)$ for any Ohnesorge number. However, according to equations \ref{eq:H_inviscid} and \ref{eq:H_viscous}, we expect different scalings for $\Deb_N \ll 1$, namely, $H/h_0 \sim \Deb_N^{2/3}$ for $\Oh \ll 1$ and $H/h_0 \sim \Deb_N$ for $\Oh \gg 1$. 

We have so far varied $\Deb$ and $\Oh$ for a fixed viscosity ratio $S=0.988$ and a fixed droplet volume characterised by a fixed value of $h_0/R_0 = 0.23$. In order to further investigate the generality of the $H/h_0$ dependence on $\Deb_N$ identified in figure \ref{fig:num_H_De}(b), we therefore performed additional simulations. Two sets of simulations were performed for $\Deb = 0.089$ and $89.2$ respectively, keeping $h_0/R_0 = 0.23$, where both $S$ and $\Oh$ were varied while keeping $S \Oh = \eta_s/\sqrt{\rho \gamma h_0}$ constant and equal to $9.88$. In each case, $S$ is varied between $0.1$ and $0.988$ where the total (constant shear) viscosity $\eta_0 = \eta_s + \eta_p$ is respectively dominated by the polymer and by the solvent contribution. All these additional data points in figure \ref{fig:num_H_De}(a) rescale on the same curve identified in figure \ref{fig:num_H_De}(b). This is because all these cases correspond to $\Oh \gg 1$ where $\tau_{\mathrm{visc}} = \eta_0 h_0 / \gamma$ is the relevant time scale of the thinning dynamics in the Newtonian regime (and not $\eta_s h_0 / \gamma$ for example), regardless of the value of $S$. 

\begin{figure}
  \centerline{\includegraphics[scale=0.58]{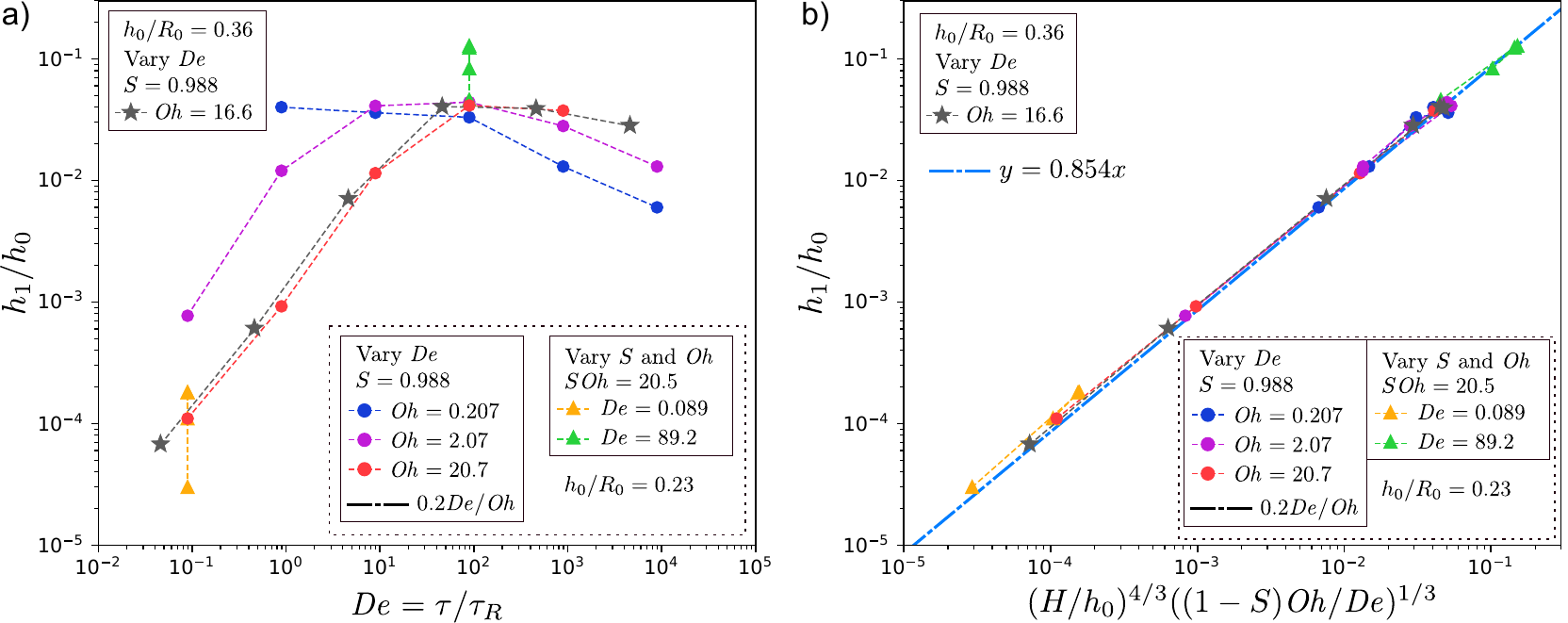}}
  \caption{Numerical (non-dimensional) transition radius $h_1/h_0$ against the Deborah number $\Deb = \tau/\tau_R$ (a) and against $(H/h_0)^{4/3} ((1-S) \Oh / \Deb)^{1/3}$ (b). The dash-dotted line in (b) is the line of equation $y = 0.854 x$. The legend and range of parameters ($\Oh$, $\Deb$, $S$ and $h_0/R_0$) are the same as in figure \ref{fig:num_H_De}.}
\label{fig:num_h1_De}
\end{figure}

Additionally, a set of simulation was performed for a larger (non-dimensional) droplet volume corresponding to $h_0/R_0 = 0.36$, varying $\Deb$ for a fixed $\Oh = 16.6$ and $S = 0.988$. These additional data points in figure \ref{fig:num_H_De}(a) also rescale on the same curve identified in figure \ref{fig:num_H_De}(b). This is because $h_0$ is the relevant time scale of the problem and sets $\tau_N$ (see equation \ref{eq:tauN}), in agreement with our experimental results which show that $h_1$ increases when increasing the droplet volume for a given plate diameter due to the increase in $h_0$ (see figure \ref{fig:h12_taue}(b)).

We can finally test the prediction of equation \ref{eq:h1_H} for the transition radius $h_1$ between the Newtonian and elastic regime which, in non-dimensional terms, reads
\begin{equation}
\frac{h_1}{h_0} = \left( \frac{H}{h_0} \right)^{4/3} \left[ \frac{(1-S)\Oh}{p \Deb} \right]^{1/3}.
\label{eq:h1_num}
\end{equation}
\noindent Figure \ref{fig:num_h1_De}(a) shows that $h_1/h_0$, which is estimated from the numerical $h/h_0$ curves for the same sets of parameters as in figure \ref{fig:num_H_De}, does not monotonically increase or decrease with $\Deb$. This is because, at low $\Deb$ (more specifically at low $\Deb_N = \tau/\tau_N$, see figure \ref{fig:num_H_De}), $H/h_0 \propto \Deb$ for $\Oh \gg 1$ according to equation \ref{eq:H_viscous}, implying that $h_1/h_0 \propto \Deb$ according to equation \ref{eq:h1_num} while, at high $\Deb$($_N$), $H/h_0 = 1$, implying that $h_1/h_0 \propto \Deb^{-1/3}$ according to equation \ref{eq:h1_num}. We find that all values of $h_1/h_0$ indeed rescale on a single master curve when potted against $(H/h_0)^{4/3} ((1-S) \Oh / \Deb)^{1/3}$ in figure \ref{fig:num_h1_De}(b) where we find that the value of $p$ in equation \ref{eq:h1_num} should be $p \approx 1.6$. Note that the value of $p$ depends on the exact definition of $h_1$, since the transition between the Newtonian and elastic regime is not necessarily sharp.

The validation of equation \ref{eq:h1_num} proves that, while equation \ref{eq:h1_norelaxation} is valid for a fast plate separation protocol, relaxation of polymer chains must be taken into account to allow for values of $H<h_0$ for a slow plate separation protocol. It also proves that, when interested in the minimum bridge / filament radius and maximum polymer extension at that point, the full 2-dimensional problem can be reduced to the simple force balance equation such as equation \ref{eq:force_balance} (which only strictly applies to a cylindrical thread) without loosing predictive power since equation \ref{eq:h1_num} (or, equivalently, equation \ref{eq:h1_H}) is based on equation \ref{eq:force_balance}.

\section{FENE-P prediction for $h_1$}
\label{sec:FENE-P prediction for $h_1$}

We now consider how finite extensibility effects, described by the FENE-P model, can affect the transition radius $h_1 = h(t_1)$ at the onset of the elastic regime. We first derive a theoretical model validated by numerical simulations in \S\ref{subsec:FENE-P Simulations} before using it in \S\ref{subsec:FENE-P Experiments} to explain some of the discrepancies observed between the Oldroyd-B theory and experiments.

\subsection{Simulations and theory}
\label{subsec:FENE-P Simulations}

\begin{figure}[t]
  \centerline{\includegraphics[scale=0.58]{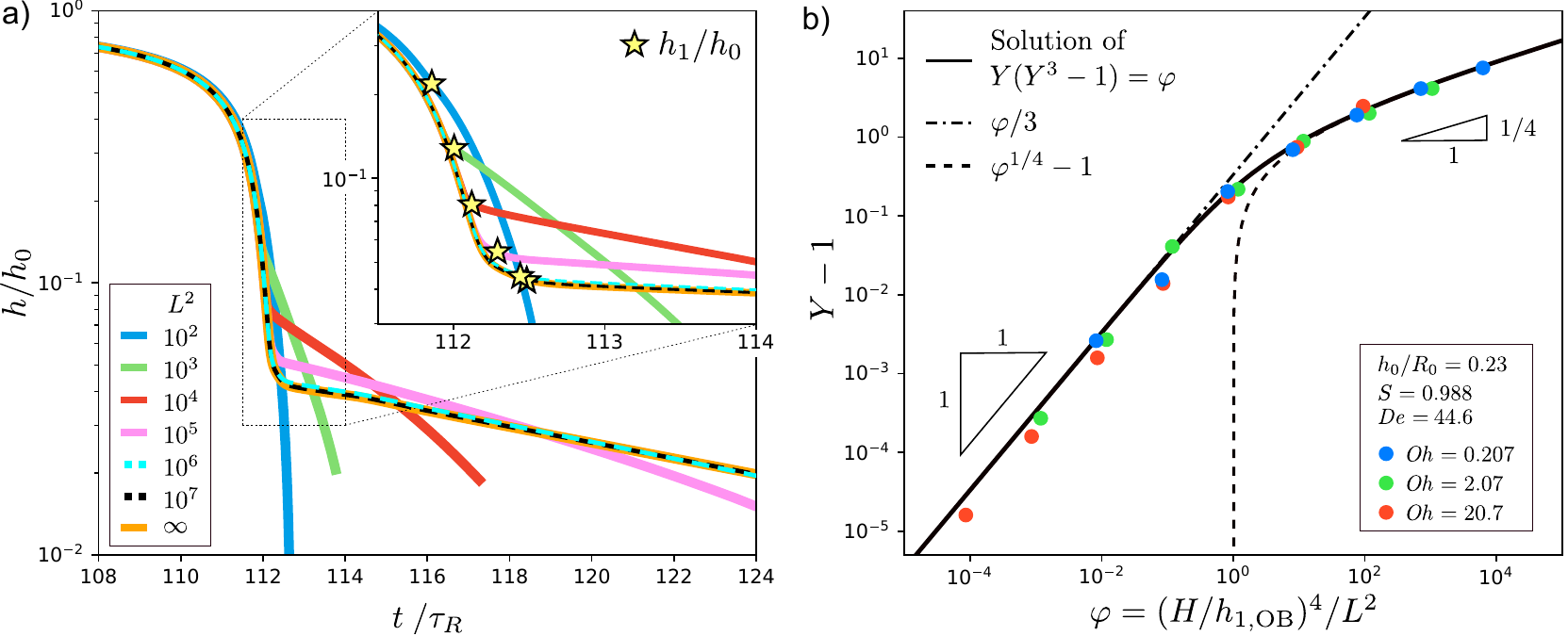}}
  \caption{(a) Numerical time evolution of $h/h_0$ using the FENE-P model with $L^2$ ranging between $10^2$ and $10^7$, as well as $L^2 = + \infty$ (Oldroyd-B limit), for fixed $\Oh = 2.07$, $\Deb = 44.6$, $S = 0.988$ and $h_0/R_0 = 0.23$. The inset figure is a zoomed version for a better visualisation of the transition to the elastic regime at $h=h_1$. Simulations start at $t=0$ used as the time reference. (b) $Y - 1$ against $\varphi$ (see equation \ref{eq:h1_FENE_2}) (where $Y = h_1 / h_{1,\mathrm{OB}}$ where $h_{1,\mathrm{OB}}$ is the Oldroyd-B limit of $h_1$) for $L^2$ ranging between $10^2$ and $10^8$ with $\Oh = 0.207$, $2.07$ and $20.7$ and fixed values of $\Deb = 44.6$, $S = 0.988$ and $h_0/R_0 = 0.23$. Values are compared with the analytical solution of equation \ref{eq:h1_FENE_2} with limit scalings of equation \ref{eq:h1_FENE_3}.}
\label{fig:h1_L2}
\end{figure}

Figure \ref{fig:h1_L2}(a) shows how, for a fixed $\Oh$, $\Deb$, $S$ and $h_0/R_0$, decreasing $L^2$ leads to an increase in $h_1$. This can be seen as counter intuitive since a decrease in $L^2$ implies that chains are shorter and therefore less elastic, which should imply a delayed transition to the elastic regime (smaller $h_1$). As we discuss in \S\ref{sec:conclusions}, this apparent contradiction is resolved by considering that shorter chains have shorter relaxation times, leading to smaller values of $H$ and therefore of $h_1$, which is not taken into account in the simulations of figure \ref{fig:h1_L2}(a) where Deborah number is kept constant. 

Note that in figure \ref{fig:h1_L2}(a), the thinning rate $\vert \dot{h}/h \vert$ in the elastic regime ($t>t_1$) is larger than $1/3\tau$ for the lowest $L^2$ values while, for the largest $L^2$ values, the elastic regime initially follows equation \ref{eq:exponential}. This is because polymer chains are already close to being fully extended at the onset of the elastic regime for the lowest $L^2$ values, as discussed in our previous paper \cite{gaillard2023beware} where we explored the possibility to invoke this effect to explain variations of the apparent relaxation time $\tau_e$ (see, e.g., figure \ref{fig:ht_water_and_PEG30}(c,e)).

The reason why $h_1$ increases as $L^2$ decreases for a fixed $\Deb$ is because the elastic stress $\sigma_{p,zz} \approx G f A_{zz}$ (assuming $A_{zz} \gg 1$ at the transition, see equation \ref{eq:miguel_sigmap}) increases faster during the Newtonian regime ($t<t_1$) as $L^2$ decreases. This is because $f \approx 1/(1-A_{zz}/L^2)$ (assuming $A_{zz} \gg 1 > A_{rr}$) diverges as $A_{zz}$ approaches $L^2$ to model the stiffening of polymer chains as they approach full extension. Formally, assuming that the transition occurs when the elastic stress reaches a fraction $p$ of the capillary pressure, i.e., when $\sigma_{p,zz} = p \gamma / h$ (where the ($2X-1$) prefactor is integrated in $p$, see \S\ref{subsec:Oldroyd-B Theory}), we get
\begin{equation}
p \frac{\gamma}{h_1} = G  \frac{A_1}{1 - A_1/L^2},
\label{eq:h1_FENE_1}
\end{equation}
\noindent where $h_1 = h(t_1)$ and $A_1 = A_{zz}(t_1)$ are the values at the transition. We assume that the bridge radius $H$ marking the onset of the coil-stretch transition is unaffected by finite extensibility effects since, at the onset of this coil-stretch transition, $A_{zz}$ is still close to $1$ and therefore $f \approx 1$. Assuming that relaxation becomes negligible between the onset of the coil-stretch transition and the onset of the elastic regime, i.e., for $h_1 < h < H$, we use $A_{zz} = (H/h)^4$ (see equation \ref{eq:Azz_4}) from which we get $A_1 = (H/h_1)^4$. Injecting this scaling in equation \ref{eq:h1_FENE_1} leads to a polynomial equation for $h_1$ in the form
\begin{equation}
h_1 = h_{1,\mathrm{OB}} \times Y(\varphi) ,
\quad Y (Y^3 - 1) = \varphi,  
\quad \varphi = \frac{A_{1,\mathrm{OB}}}{L^2}, 
\label{eq:h1_FENE_2}
\end{equation}
\noindent where
\begin{equation}
h_{1,\mathrm{OB}} = \left( \frac{G H^4}{p \gamma} \right)^{1/3} 
\quad \mathrm{and} \quad A_{1,\mathrm{OB}} = \left( \frac{H}{h_{1,\mathrm{OB}}} \right)^4 
\label{eq:h1OB}
\end{equation}
\noindent are the value of $h_1$ and $A_1$ predicted in the Oldroyd-B limit $L^2 = + \infty$ ($\varphi = 0$), see equation \ref{eq:h1_H} for $h_{1,\mathrm{OB}}$. The two limit scalings of equation \ref{eq:h1_FENE_2} corresponding to weak ($\varphi \ll 1$) and strong ($\varphi \gg 1$) finite extensibility effects are
\begin{equation}
Y = 
\begin{cases} 
1 + \varphi/3   & \text{for } \varphi \ll 1, \\
\varphi^{1/4} & \text{for } \varphi \gg 1 .
\end{cases}
\label{eq:h1_FENE_3}
\end{equation}
\noindent In the weak limit ($\varphi \ll 1$), polymer chains are still far from full extension at the onset of the elastic regime while, in the strong limit ($\varphi \gg 1$), the transition occurs significantly sooner than the Oldroyd-B prediction due to the fact that polymer chain have almost already reached full extension at the transition. Indeed, in the strong limit, the transition occurs when $A_{zz} = (H/h)^4$ becomes of the order of $L^2$, leading to 
\begin{equation}
h_1 \sim \frac{H}{L^{1/2}},
\label{eq:h1_lowL2}
\end{equation}
\noindent which is equivalent to equation \ref{eq:h1_FENE_3} for $\varphi \gg 1$.

Values of $Y$ estimated from numerical simulations using the FENE-P model are plotted against $\varphi$ in figure \ref{fig:h1_L2}(b) for three different Ohnesorge numbers, varying $L^2$ between $10^2$ and $10^8$ in each case while keeping $\Deb$, $S$ and $h_0/R_0$ constant, using an extra Oldroyd-B simulation as a reference to get $h_{1,\mathrm{OB}}$. We find that all data points collapse on a single curve corresponding to the solution of equation \ref{eq:h1_FENE_2}. In particular, no prefactor is required in the $\varphi \gg 1$ limit. Note that we choose to plot $Y - 1$ instead of $Y$ in order to better visualise the $\varphi \ll 1$ regime. Note that the values of $H/h_0$, estimated for each simulation like in figure \ref{fig:num_h_Azz}, is found not to depend on $L^2$ and takes values between $1$ and $0.36$ for the data of figure \ref{fig:h1_L2}(b). Equation \ref{eq:h1_FENE_2} therefore generalises equation \ref{eq:h1_norelaxation} to cases where both finite extensibility effects and polymer relaxation effects are not negligible.

\subsection{Experiments}
\label{subsec:FENE-P Experiments}

Experimentally, cases where both polymer relaxation and finite extensibility effects are expected to play a role correspond to low polymer concentrations $c$. Indeed, as $c$ decreases, the transition to the elastic regime is delayed (i.e. $h_1$ decreases, see figures \ref{fig:ht_varyc}(b) and \ref{fig:h012_tau_varyc}(b)) and polymer chains are therefore expected to be increasingly stretched at the onset of the elastic regime, assuming that their relaxation time becomes equal to the Zimm relaxation $\tau_Z$ time which is independent of polymer concentration $c$ (see equation \ref{eq:zimm}). Polymer chains should therefore ultimately approach full extension (at the onset of the elastic regime) below a critical concentration $c_{\mathrm{low}}$ introduced by Campo-Deaño \& Clasen \cite{campo2010slow}. For $c<c_{\mathrm{low}}$, the elasto-capillary balance leading to the exponential decay of equation \ref{eq:exponential} is therefore no longer valid since $A_1 \sim L^2$, leading to filaments thinning rates $\vert \dot{h}/h \vert > 1/3\tau$, as shown numerically in figure \ref{fig:h1_L2}(a) and discussed in our previous paper \cite{gaillard2023beware}. This would explain why fitting the elastic regime ($t>t_1$) with a exponential leads to apparent relaxation times $\tau_e$ that are smaller than $\tau_Z$, as reported in figure \ref{fig:h012_tau_varyc}(a) for aqueous PEO-4M solutions of concentrations $c \le 10$~ppm. This would also explain why, as discussed in \S\ref{subsec:Oldroyd-B Experiments} (see triangle symbols in figure \ref{fig:GH}(b)), transition radii $h_1$ measured for low polymer concentrations cannot be captured by the Oldroyd-B prediction even when replacing values of $\tau_e<\tau_Z$ by $\tau_Z$.

\begin{figure}[t]
  \centerline{\includegraphics[scale=0.58]{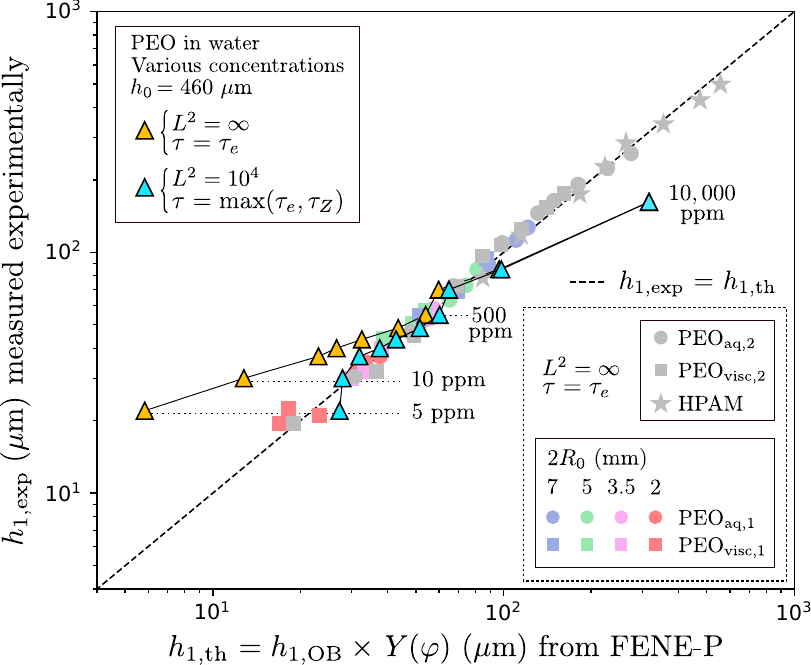}}
  \caption{Experimentally measured $h_1$ values ($h_{1,\mathrm{exp}}$) against FENE-P's theoretical prediction $h_{1,\mathrm{th}} = h_{1,\mathrm{OB}} \times Y(\varphi)$ (see equations \ref{eq:h1_FENE_2} and \ref{eq:h1OB}) for various polymer solutions and initial bridge radii. Values of the FENE-P model parameters $\tau$ and $L^2$ are indicated in legends ($\eta_p$ values are the ones from shear rheology). The discrepancy between experiments and the Oldroyd-B prediction for $h_1$ at low polymer concentration can be rationalised by finite extensibility effects, see main text.}
\label{fig:h1_fenep_lowc}
\end{figure}

This idea is tested in figure \ref{fig:h1_fenep_lowc} where experimentally measured values of $h_1$ ($h_{1,\mathrm{exp}}$) are plotted against the FENE-P theoretical prediction $h_{1,\mathrm{th}} = h_{1,\mathrm{OB}} \times Y(\varphi)$ (see equations \ref{eq:h1_FENE_2} and \ref{eq:h1OB}) for all polymer solutions and initial bridge radii. We choose $L^2 = \infty$ ($Y = 1$) and $\tau = \tau_e$ as model parameters for data points corresponding to the PEO$_{\mathrm{aq}}$ (1 and 2, $500$~ppm PEO-4M solution in water), PEO$_{\mathrm{visc}}$ (1 and 2, $25$~ppm PEO-4M solution in a more viscous solvent) and HPAM solutions since we already know from figure \ref{fig:GH}(b) that these $h_1$ values are consistent with the Oldroyd-B prediction of equation \ref{eq:h1_H} (we choose $p = 0.27$ to match the prefactor found in figure \ref{fig:GH}(b)). The experimentally measured values of $h_1$ corresponding aqueous PEO-4M solutions of various concentrations in figure \ref{fig:h1_fenep_lowc} (orange triangle symbols) are higher that the Oldroyd-B prediction at low concentrations, as we saw in figure \ref{fig:GH}(b) where replacing values of $\tau_e < \tau_Z$ by $\tau_Z$ was found to be insufficient to explain the discrepancy. The blue triangle symbols in figure \ref{fig:h1_fenep_lowc} show that this discrepancy can be rationalised using the FENE-P model where we chose a value of $L^2 = 1 \times 10^4$ sufficiently small to allow for values of $Y$ sufficiently larger than $1$ (i.e., polymer chains close to being fully extended at the onset of the elastic regime) to ``fill the remaining gap'', while using $\tau = \max(\tau_e,\tau_Z)$ at the model relaxation time.

This value of $L^2 = 1 \times 10^4$ has the same order of magnitude as the value expected from the microscopic formula \cite{clasen2006dilute}
\begin{equation}
L^2 = 3 \left[ \frac{j \sin^2{(\theta/2)} M_w}{C_{\infty} \, M_u} \right]^{2(1-\nu)} \,\mathrm{,}
\label{eq:L2_microscopic}
\end{equation}
\noindent which gives $L^2$ between $1.6 \times 10^{4}$ and $1.3 \times 10^{5}$ for PEO of molecular weight $M_w = 4 \times 10^{6}$~g/mol, for solvent quality exponents $\nu$ between $3/5$ (good solvent) and $1/2$ (theta solvent), where $M_u$ is the monomer molecular weight, $\theta = 109^{\circ}$ the C-C bond angle, $j=3$ the number of bonds of a monomer and $C_{\infty} = 4.1$ the characteristic ratio \citep{brandrup1989polymer}. Assuming that $\nu$ is between $1/2$ and $3/5$, the discrepancy can be explained by a value $M_w$ less than $4 \times 10^{6}$~g/mol stemming from polymer degradation during mixing when preparing the stock solution. Note that choosing $L^2 = 1 \times 10^4$ leads to values of $Y$ close to $1$ (up to $1.18$) for the PEO$_{\mathrm{aq}}$ (1 and 2), PEO$_{\mathrm{visc}}$ (1 and 2) and HPAM solutions, consistent with the agreement between experiments and the Oldroyd-B theory for these solutions in figure \ref{fig:h1_fenep_lowc}, as shown by the small difference between the $L^2 =\infty$ (orange) and $L^2 = 1 \times 10^4$ (blue) triangle symbols for $c = 500$~ppm which corresponds to the PEO$_{\mathrm{aq,1}}$ solution.

%%%%%%%%%%%%%%%%%%%%%%%%%%%%%%%%%%%%%%%%%%%%%%%%%%%%%%%%%%%%%%%%%%%%%%%%%%%%%%%%%%%%%%%%%%%%%%%%%%%%%%%%%%%%%%%%%%%%%%%%%%%%%%%%
\section{Conclusions and discussions}
\label{sec:conclusions}

We have shown experimentally that the classical formula of equation \ref{eq:h1_norelaxation} for the bridge / filament radius $h_1$ marking the onset of the elastic regime does not hold for slow filament thinning techniques such as CaBER with a slow plate separation protocol. This is because, unlike what is assumed to derive equation \ref{eq:h1_norelaxation}, polymer chains do not necessarily start stretching (beyond their equilibrium coiled configuration) at the threshold of the Rayleigh-Plateau instability (with minimum bridge radius $h_0$), but only do so when the bridge thinning rate becomes comparable to the inverse of their relaxation time. This was confirmed numerically using the Oldroyd-B model where we have shown that, in the low-relaxation-time limit where polymer chains only start stretching when the bridge thinning dynamics has become self-similar, $h_1$ does not depend on $h_0$ anymore, as anticipated by Campo-Deaño \& Clasen \cite{campo2010slow}. Our generalised formula (see equation \ref{eq:h1_H} or \ref{eq:h1_num}) is in principle valid for any pinch-off experiment, e.g., in dripping experiments, provided that polymer chains are relaxed at the onset of capillary thinning (no pre-stress). 

This formula was extended to finitely extensible polymer chains using a FENE-P description, which solved the observed discrepancy between the Oldroyd-B theory and values of $h_1$ measured for low polymer concentrations. This is consistent with the fact that apparent relaxation times $\tau_e$ less than the Zimm relaxation times were measured at these low concentrations, an anomaly also reported by Campo-Deaño \& Clasen \cite{campo2010slow} which stems from polymer chains being close to full extension at the onset of the elastic regime, a case where equation \ref{eq:exponential} is no longer valid and thinning rates $\vert \dot{h}/h \vert > 1/3\tau$ are observed, as we discuss in our previous paper \cite{gaillard2023beware}.

Our generalised Oldroyd-B formula of equation \ref{eq:h1_H} solves an apparent paradox of equation \ref{eq:h1_norelaxation} which predicts that the transition to the elastic regime occurs sooner as polymer chains get shorter for a fixed (mass) concentration. Indeed, equation \ref{eq:h1_norelaxation} predicts $h_1 \propto G^{1/3}$ where the elastic modulus scales with molecular weight as $G \propto M_w^{-1}$ in the Rouse-Zimm theory \cite{clasen2006dilute}, yielding $h_1 \propto M_w^{-1/3}$ which increases as $M_w$ decreases. This is counter intuitive since shorter chains should imply lower elasticity and therefore a delayed transition, $h_1$ approaching $0$ as polymer chains approach monomer size. This apparent paradox is solved by realising that shorter chains have a shorter (Zimm) relaxation time since $\tau_Z \propto M_w^{3\nu}$ \cite{clasen2006dilute}, where $\nu$ is the solvent quality exponent, implying that $H$ (see equation \ref{eq:h1_H}) should start decreasing as $M_w$ decreases for sufficiently low Deborah numbers (see figure \ref{fig:num_H_De}). In the low-relaxation-time (equivalently low-$M_w$) limit where polymer chains only start stretching when the bridge thinning dynamics has become self-similar, we get $h_1 \propto M_w^{(8\nu -1)/3}$ for $\Oh \ll 1$ and $h_1 \propto M_w^{(12\nu -1)/3}$ for $\Oh \gg 1$ according to equations \ref{eq:h1_inviscid} and \ref{eq:h1_viscous}, both exponents being positive for any $\nu$ between $1/2$ (theta solvent) and $3/5$ (good solvent). This remains true even in the limit where polymer chains are almost fully extended at the onset of the elastic regime since equation \ref{eq:h1_lowL2} predicts $h_1 \propto M_w^{(5\nu -1)/2}$ for $\Oh \ll 1$ and $h_1 \propto M_w^{(7\nu -1)/2}$ for $\Oh \gg 1$ according to equations \ref{eq:H_inviscid} and \ref{eq:H_viscous} for $H$, using $L \propto M_w^{1-\nu}$ (see equation \ref{eq:L2_microscopic}).

We saw in \S\ref{subsec:Oldroyd-B Experiments} that experimental values of $h_1$ can only be rationalised using the apparent ($h_0$-dependent) relaxation time $\tau_e$ as `the' relaxation time, in contradiction with the idea discussed in our previous paper \cite{gaillard2023beware} that the `real' relaxation time should be high-$h_0$ limit value $\tau_m$. Indeed, when choosing $\tau_m$, equation \ref{eq:h1_H} only works for the PEO$_{\mathrm{visc}}$ solution, i.e., the most dilute one in the most viscous solvent. This would suggest that $\tau_e$ is the `true' relaxation time, and not $\tau_m$. This implies that a given polymer solution should exhibit the same thinning rate regardless of the system size (plate diameter or droplet volume), inconsistent with our observations of a system-size dependent apparent relaxation time $\tau_e$ (see figure \ref{fig:ht_water_and_PEG30}(c,e)). Therefore, if $\tau_e$ really measures the `true' relaxation, it implies that some rheological property of a polymer solution somehow `changes' when being tested with a different plate diameter via a mechanism which we could not identify and which is unlikely to be evaporation or polymer degradation, as we discussed in our previous paper \cite{gaillard2023beware}. Another possibility is that the solution in fact does not change, meaning that the system-size dependence of $\tau_e$ is \emph{not} an artefact, in which case it would be only by coincidence that we could successfully capture experimental values of $h_1$ using $\tau_e$. This would imply that the Oldroyd-B and FENE-P models miss some important features of polymer dynamics in extensional flows, strengthening the already established need for better constitutive equations. Future works will determine if more sophisticated models such as Conformation-Dependent Drag (CDD) models, accounting for the action of both chain stretching and intermolecular hydrodynamic interactions on the friction coefficient \cite{prabhakar2016influence,prabhakar2017effect}, are able to rationalise our experimental results on the system-size dependence of both $\tau_e$ and $h_1$. \\

\textbf{Conflicts of interest} The authors declare no conflicts of interest.\\

\textbf{Funding} M. A. Herrada acknowledges funding from the Spanish Ministry of Economy, Industry and Competitiveness under Grant PID2022-140951O.

%\textbf{Acknowledgments} 

\begin{small}
\bibliographystyle{elsarticle-num} 
\bibliography{bibliography.bib}
\end{small}

%\begin{thebibliography}{00}


%\bibitem{}

%\end{thebibliography}
\end{document}